\documentclass[AMA,STIX1COL,doublespace]{WileyNJD-v2}
\usepackage{moreverb}
\usepackage{bui}
\usepackage{lineno}
\usepackage{stix}
\usepackage{hyperref}
\hypersetup{
    colorlinks,
    citecolor=blue,
    filecolor=black,
    linkcolor=blue,
    urlcolor=blue
}
\usepackage{cleveref}
\usepackage{csquotes}

\newcommand\BibTeX{{\rmfamily B\kern-.05em \textsc{i\kern-.025em b}\kern-.08em
T\kern-.1667em\lower.7ex\hbox{E}\kern-.125emX}}
\renewcommand{\gm}[1]{\textcolor{black}{#1}}

\articletype{Journal Article}%

\received{<day> <Month>, <year>}
\revised{<day> <Month>, <year>}
\accepted{<day> <Month>, <year>}

\begin{document}

%\title{Implicit Integration with Sub-stepping Scheme for Critical State Models}
\title{Implicit Sub-stepping Scheme for Critical State Soil Models}

\author[1]{Hoang-Giang Bui}

\author[1]{Jelena Nin\'ic}

\author[2]{G\"unther Meschke}

\authormark{BUI \textsc{et al}}

\address[1]{\orgdiv{University of Birmingham}, \orgname{School of Engineering}, \orgaddress{\state{Birmingham}, \country{UK}}}

\address[2]{\orgdiv{Ruhr University Bochum}, \orgname{Institute for Structural Mechanics}, \orgaddress{\state{Bochum}, \country{Germany}}}

%\corres{*G\"unther Meschke, Universit\"atstr 150, 44801, Bochum. \email{guenther.meschke@rub.de}}
\corres{Hoang Giang Bui, University of Birmingham, UK. \email{b.hoanggiang@bham.ac.uk}}

%\presentaddress{Universit\"atstr 150, 44801, Bochum, Germany}

%\abstract[Abstract]{
%The stress integration of critical soil model is usually based on implicit Euler algorithm, where the stress predictor is corrected by employing a return mapping algorithm. The solution on local nonlinear system is sometimes not achieved, due to large loading step. To overcome this problem, a sub-stepping scheme is normally used, to improve the convergence of the local nonlinear system. Nevertheless, the complexity of the tangent operator of the sub-stepping scheme arises enormously, if more sub-steps are used in the integration. It inhibits the use of the implicit sub-stepping scheme, to maintain the quadratic rate of convergence of the global Newton-Raphson algorithm. This paper addresses this issue by deriving an explicit equation for implicit sub-stepping stress integration and consistent linearization for the modified Cam-Clay model and unified Clay and Sand model.
%}

\abstract[Abstract]{
The stress integration of critical soil model is usually based on implicit Euler algorithm, where the stress predictor is corrected by employing a return mapping algorithm. In the case of large load step, the solution of local nonlinear system to compute the plastic multiplier may not be attained. To overcome this problem, a sub-stepping scheme shall be used to improve the convergence of the local nonlinear system solution strategy. Nevertheless, the complexity of the tangent operator of the sub-stepping scheme is high. This complicates the use of Newton-Raphson algorithm to obtain global quadratic convergence. In this paper, a formulation for consistent tangent operator is developed for implicit sub-stepping integration for the modified Cam-Clay model and unified Clay and Sand model. This formulation is highly efficient and can be used with problem involving large load step, such as tunnel simulation.
}


\jnlcitation{\cname{%
\author{Hoang-Giang Bui}, 
\author{G\"unther Meschke}} (\cyear{2020}), 
\ctitle{Implicit Integration with Sub-stepping Scheme for Critical State Models}, \cjournal{International Journal for Numerical Methods in Engineering}, \cvol{xxxx;xx:xx--xx}.}

\maketitle

%\footnotetext{\textbf{Abbreviations:} ANA, anti-nuclear antibodies; APC, antigen-presenting cells; IRF, interferon regulatory factor}

%\linenumbers

\section{Introduction}
\label{sec:intro}
%
% introduce critical state soil model
% introduce cam-clay and modified cam-clay, latest development
% introduce CASM
%
The introduction of critical soil models marks an important advance in computational geomechanics. Based on elastoplastic theory, the model is first proposed by Roscoe and coworkers \cite{Roscoe_Schofield:63,Schofield_Wroth:68} as Cam-Clay model (CC), and subsequently improved as modified Cam-Clay model \cite{Roscoe_Burland:68}. The modified Cam-Clay model (MCC) is shown to sufficiently predict \gm{main attributes of the constitutive behavior of normally consolidated soft cohesive soils} \citep{Gens_Potts:88}, \gm{although deficiencies were reported when modeling highly overconsolidated soils}.  \gm{Later,} Yu \cite{Yu:98}  generalized the standard Cam-Clay model to \gm{allow  modeling} the behavior of sand. The unified model, \gm{denoted as} CASM \cite{Yu:98,Yu_Khong_Wang:07}, has the ability to resolve some drawbacks of the standard Cam-Clay models, such as the overestimation of \gm{the strength} on the \textquote{dry side} or the behavior of loose sand in undrained tests, while \gm{limiting} the number of parameters to seven, which is an advantage to adopt the model both in academic \gm{research}  and practical applications.

To apply the critical soil models in numerical simulations, an integration algorithm is required. In the context of implicit methods, the backward Euler algorithm is typically used. This is frequently resorted to the return mapping algorithm \cite{Simo_Hughes:98,Simo_Taylor:86}, which will ensure quadratic convergence of the Newton-Raphson algorithm if certain conditions are satisfied. The advantage of the implicit method is its robustness, second-order accuracy and able to use with large load step. Nevertheless, the implicit algorithm typically requires the high-order derivatives of yield functions and plastic flow rule. The return mapping algorithm \gm{has been} successfully applied to Cam-Clay models in \cite{Borja:90} and, being extended to the finite strain regime, in \cite{Simo_Meschke:93}.
\gm{Further applications of the return map algorithm to extensions of the Cam Clay model have been suggested in \cite{Borja_Lin_Montans:01,Borja_Choo:16,Borja:04}.}

In contrast to implicit methods, explicit methods, based on the forward Euler integration, requires only the first order approximation of the yield surface and plastic flow rule and it is considered to be more straightforward to implement. Because only first order accuracy is attained, the load step must be sufficiently small to reduce the accumulated error. The efficiency and robustness of explicit algorithms has been \gm{substantially}  improved by using sub-stepping algorithms \cite{Sloan:87,Sloan_Abbo_Sheng:01,Sheng_Sloan:01,Zhao_et_al:05}.
%\gm{*** GM *** CHECK IF PREVIOUS ARE WITHOUT ERROR CONTROL ***
\gm{An automatic error control for sub-stepping algorithm is proposed in \cite{Abbo_Sloan:96}.}

 \gm{It is well known from analysis of convergence properties of stability of the Newton algorithm, from applications reported in the literature \cite{Crisfield:91,Borja:13}, and also experienced by the authors, that the computational robustness of implicit integration schemes applied to elastoplastic models, and, in particular, to critical state models, is limited, and that no convergence of the local iterative scheme at the integration point level is attained when the load step is large.}
%As observed by the authors, the failure of the stress integration of the critical state model often occurs in solving the local nonlinear system stemming from the yield condition and plastic flow rule, rather than failure of the global Newton-Raphson loop. 
 Therefore, it is crucial to sub-divide the load into smaller increments to ensure the existence of local admissible stress state and the convergence of the local iteration, \gm{to} compute the internal variables and \gm{the} plastic increment. Because the sub-load increment can be adjusted, this scheme is highly effective for problems, \gm{in} which the loading path is difficult to control. \gm{An example of this class of problems are excavation problems, which are a priori stress control problems if no arc length methods are applied.} 
 \gm{A consistent sub-stepping scheme has been proposed for elastoplastic models in \citep{Perez-Fouget_Rodriguez-Ferran_Huerta:01}. In this paper the trial stress predictor uses a linear extrapolation based on a (constant) elastic tangent, which is typically used for elastoplastic material models. However, } this scheme is not optimal for critical state soil models, because the stress-strain rule in the elastic region of the critical state model is nonlinear \cite{Roscoe_Burland:68}. 
 
 In \gm{\Cref{sec:mcc} of this paper, a sub-stepping algorithm for the implicit integration algorithm for critical state soil models, more specifically for the Modified Cam Clay model and the Clay and Sand model, is proposed,  which makes use of the more accurate nonlinear approximation of hydrostatic pressure. The effectiveness of the implicit sub-stepping scheme is demonstrated} by selected numerical examples presented in \Cref{sec:numerical_examples}. \gm{The findings from these analyses are concluded in \Cref{sec:concl}.}

\section{Implicit integration of the critical state soil models}
\label{sec:mcc}
 In this section,  the basic governing equations of the Modified Cam-Clay (MCC) (\Cref{sec:cam-clay-overview}) model and the Clay and Soil model (\Cref{sec:clay-sand-overview})  are summarized before an implicit integration algorithm incorporating sub-stepping is elaborated for both models in subsections \Cref{sec:mcc_multi}.
\subsection{Brief summary of the modified Cam-Clay model}
\label{sec:cam-clay-overview}
The \gm{Modified Cam-Clay (MCC) model} \cite{Schofield_Wroth:68,Roscoe_Schofield_Wroth:58} is characterized by \gm{a} yield surface \gm{formulated in the} $p-q$ plane by:
\begin{equation}
f^{\text{MCC}}(p, q, p_c) = \left( \dfrac{q}{M} \right)^2 + p (p - p_c),
\label{eq:yield_mcc}
\end{equation}
\gm{by the associative plastic potential $g^{\text{MCC}}(p, q, p_c) = f^{\text{MCC}}(p, q, p_c)$ and by the nonlinear stress-strain relation in the elastic domain:}
\begin{equation}
\dot{p} = K \dot{\varepsilon}^e_v \, ,
\qquad
\dot{\mathbf{s}} = 2 \mu \dot{\bs{\varepsilon}}^e_d .
\label{eq:elastic_mcc}
\end{equation}
The \gm{shear modulus and the pressure-dependent compression modulus $K$ are given as}
\begin{equation}
K = \dfrac{1 + e}{\kappa} p \, ,
\qquad
\mu=Kr \, ,
\qquad
r = \dfrac{3}{2} \dfrac{1-2\nu}{1+\nu},
\label{eq:stiff_mcc}
\end{equation}
\gm{and the hardening rule is specified as}
\begin{equation}
\dot{p}_c = \theta p_c \dot{\varepsilon}^p_v \, ,
\qquad
\theta = \dfrac{1+e}{\lambda-\kappa} .
\label{eq:hardening_mcc}
\end{equation}
In \Cref{eq:yield_mcc,eq:elastic_mcc,eq:stiff_mcc,eq:hardening_mcc}, $M, e, \lambda, \kappa, \nu$ are material constants and have following meanings: $M$ is the slope of the critical state line, $e$ is the void ratio, $\lambda$ is the compression index, $\kappa$ is the swelling index and $\nu$ is the Poisson ratio. For the details on the material parameters and their roles in reflecting the characteristics of soil, the reader is referred to \cite{Wood:90}. $p$ denotes the hydrostatic pressure and $q$ is the deviatoric invariant $q=\sqrt{3/2 \, \bf{s}:\bf{s}}$, with 
$\mathbf{s} = \bs{\sigma} - p \mathbf{I}$
as the deviatoric stress and $p$= $\dfrac{1}{3} \text{tr} (\bs{\sigma})$. The subscript $(\bullet)_v$ denotes the volumetric component and $(\text{•})_d$ denotes the deviatoric component of the strain. The superscript $(\bullet)^e$ denotes the elastic component and $(\bullet)^p$ denotes the plastic component of the strain. The void ratio $e$ is considered not to change significantly during a load step and is kept fixed \gm{within one increment}.

\subsection{Brief summary of the CASM model}
\label{sec:clay-sand-overview}
The \gm{Clay and Sand model (CASM)}  model, as proposed in \cite{Yu:98}, adopts the yield surface:
\begin{equation}
f^{\text{CASM}}(p, q, p_c) = \left( \dfrac{q}{Mp} \right)^N + \dfrac{1}{\ln R} \ln \dfrac{p}{p_c} ,
\label{eq:yield_casm}
\end{equation}
and the non-associative plastic potential:
\begin{equation}
g^{\text{CASM}}(p, q) = 3M \ln \dfrac{p}{\beta} + (3+2M) \ln \left( 2 \dfrac{q}{p} + 3 \right) - (3-M) \ln \left( 3 - \dfrac{q}{p} \right) .
\label{eq:casm_plastic_potential}
\end{equation}
$\beta$ is a  size parameter and dependent on the stress state. It can be obtained by solving $g^{\text{CASM}}(p, q) = 0$ at a plastic state. 
The stress-strain relation in the elastic domain and the hardening rule \eqref{eq:hardening_mcc} are similar to the MCC model (see \eqref{eq:elastic_mcc} and \eqref{eq:stiff_mcc}). In addition to the material parameters $M$, $e$, $\lambda$, $\kappa$, $\mu$, the CASM model is characterized by additional parameters \gm{$N$ and $R$ defining the shape of the yield surface in the deviatoric plane.}  $N$ is the parameter controlling the shape of the yield surface and $R$ is the spacing ratio. 
The CASM model can be adjusted to reproduce the standard Cam-Clay model. For example, the original CC model can be obtained by setting $N=1$ and $\ln R=1$, while the MCC model can be approximated by setting $R=2$ and a suitable value for $N$. For more details on the theoretical basis of CASM model, the reader is referred to \cite{Yu:98}.
\subsection{One-step implicit integration scheme}
\label{sec:mcc_one}
In \gm{the one-step integration scheme}, the hydrostatic pressure is approximated by integrating the rate form of nonlinear stress-strain relation (see \Cref{eq:elastic_mcc}) \cite{Borja:91a}. It is summarized here to serve as the basis for sub-stepping scheme proposed in the next Subsection.
\subsubsection{Stress integration algorithm}
\label{sec:mcc_stress_int}
We denote $\{p_{n+1}, q_{n+1}, (p_c)_{n+1}\}$ as
the state variables at the current load step and $\Delta \boldsymbol{\varepsilon} = \boldsymbol{\varepsilon}_{n+1} - \boldsymbol{\varepsilon}_n$ as the incremental total strain obtained from the solution of the equilibrium equations on the structural level for load step $n+1$.   For brevity, the step index $n+1$ of the current step is neglected in what follows. In the predictor step, which assumes pure elastic loading ($\Delta \varepsilon^e_v = \Delta \epsilon_v$), the stresses and shear stiffness are calculated as:
\begin{equation}
p = p_n \, \text{exp} \left( \dfrac{1+e}{\kappa} \Delta \epsilon_v \right) \, ,
\qquad
\mathbf{s} = \mathbf{s}_n + 2 \bar{\mu} \Delta \bs{\varepsilon}_d \, ,
\qquad
p_c = (p_c)_n \, ,
\qquad
\bar{\mu} = \dfrac{p - p_n}{\Delta \epsilon_v} r .
\label{eq:mcc_elastic_stresses}
\end{equation}

If the yield condition in the predictor step is violated ($f(p, q, p_c) > 0$), the 
material point is subjected to plastic loading, \gm{and $\{p,q,p_c\}$ along with the increment of the plastic multiplier $\Delta \phi$ need to be determined from the solution of the following set of nonlinear equations, which holds for any Critical State model:} 
\begin{align}
h_1 &= p - p_n \, \text{exp} \left( \dfrac{1+e}{\kappa} \Delta \varepsilon^e_v \right) = 0, \label{eq:p_mcc} \\
h_2 &= q - \sqrt{\dfrac{3}{2}} \| \mathbf{s}_n + 2\bar{\mu} \Delta \boldsymbol{\varepsilon}_d^e \| = 0 \, , \quad \bar{\mu} = \dfrac{p - p_n}{\Delta \epsilon_v^e} r, \label{eq:q_mcc} \\
h_3 &= p_c - \left( p_c \right)_n \text{exp} \left( \theta \Delta \epsilon_v^p \right) = 0, \label{eq:pc_mcc} \\
h_4 &= f(p, q, p_c) = 0 . \label{eq:f_mcc}
\end{align}
%
%%TODO add the appendix to prove the existence of the solution of the local nonlinear system
%
The \gm{solution for the set of unknowns}  $\{ p, q, p_c, \Delta \phi \}$ of the nonlinear system \cref{eq:p_mcc,eq:q_mcc,eq:pc_mcc,eq:f_mcc} can be determined by using the implicit Newton-Raphson algorithm. For details of the linearization of the \gm{residuals from the  nonlinear system when using the Newton-Raphson method for the} MCC model, the reader is referred to \cite{Borja:91a}. For the CASM model, the reader is referred to \Cref{app:casm_local_solution}.

\subsubsection{Consistent tangent operator}
\label{sec:mcc_tangent}
With the \gm{solution for the stress state and the hardening variables} at hand, the consistent tangent operator can be defined as:
\begin{equation}
\mathbb{C} = \dfrac{\partial \boldsymbol{\sigma}
}{\partial \Delta \boldsymbol{\varepsilon}} = \dfrac{\partial \mathbf{s}}{\partial \Delta \boldsymbol{\varepsilon}} + \mathbf{I} \otimes \dfrac{\partial p}{\partial \Delta \boldsymbol{\varepsilon}} .
\label{eq:tangent_operator}
\end{equation}
The consistent tangent operator $\mathbb{C}$ is crucial to preserve the quadratic rate of convergence of the fully implicit Newton-Raphson strategy. In the elastic state, the components $\dfrac{\partial \mathbf{s}}{\partial \Delta \boldsymbol{\varepsilon}}$ and $\dfrac{\partial p}{\partial \Delta \boldsymbol{\varepsilon}}$ of \eqref{eq:tangent_operator} can be straightforwardly calculated by taking \gm{the} linearization of \eqref{eq:mcc_elastic_stresses}:
\begin{equation}
\dfrac{\partial p}{\partial \Delta \boldsymbol{\varepsilon}} = K \mathbf{I} \, ,
\quad
\dfrac{\partial \mathbf{s}}{\partial \Delta \boldsymbol{\varepsilon}} = 2 \bar{\mu} \mathbb{I}_d + 2 \Delta \bs{\varepsilon}_d \otimes \dfrac{\partial \bar{\mu}}{\partial \Delta \boldsymbol{\varepsilon}} \, ,
\quad
\dfrac{\partial \bar{\mu}}{\partial \Delta \boldsymbol{\varepsilon}} = \dfrac{K r - \bar{\mu}}{\Delta \epsilon_v} \mathbf{I} \, ,
\quad
K = \dfrac{1+e}{\kappa} p .
\end{equation}

In the case that the \gm{strain increment in the integration point} is \gm{leading to an elasto-plastic response in the current increment}, $\dfrac{\partial p}{\partial \Delta \boldsymbol{\varepsilon}}$, $\dfrac{\partial p_c}{\partial \Delta \boldsymbol{\varepsilon}}$ and $\dfrac{\partial \mathbf{s}}{\partial \Delta \boldsymbol{\varepsilon}}$ \gm{can be written in terms} of $\dfrac{\partial \Delta \phi}{\partial \Delta \boldsymbol{\varepsilon}}$ as:
\begin{equation}
\dfrac{\partial p}{\partial \Delta \boldsymbol{\varepsilon}} = \mathbf{A} + a \dfrac{\partial \Delta \phi}{\partial \Delta \boldsymbol{\varepsilon}} \, ,
\qquad
\dfrac{\partial p_c}{\partial \Delta \boldsymbol{\varepsilon}} = \mathbf{B} + b \dfrac{\partial \Delta \phi}{\partial \Delta \boldsymbol{\varepsilon}} \, ,
\qquad
\dfrac{\partial \mathbf{s}}{\partial \Delta \boldsymbol{\varepsilon}} = \mathbb{D} + \mathbf{D} \otimes \dfrac{\partial \Delta \phi}{\partial \Delta \boldsymbol{\varepsilon}} .
\label{eq:mcc_p_de_and_s_de}
\end{equation}
\gm{In analogy, linearization of $q$ is obtained as:}
\begin{equation}
\dfrac{\partial q}{\partial \Delta \boldsymbol{\varepsilon}} = \dfrac{3}{2} \dfrac{1}{q} \left( \mathbf{s} : \dfrac{\partial \mathbf{s}}{\partial \Delta \boldsymbol{\varepsilon}} \right) = \dfrac{3}{2} \dfrac{1}{q} \left( \mathbf{s} : \mathbb{D} + \left( \mathbf{s} : \mathbf{D} \right) \dfrac{\partial \Delta \phi}{\partial \Delta \boldsymbol{\varepsilon}} \right).
\label{eq:mcc_pc_de_and_q_de}
\end{equation}

For \gm{the} MCC model, linearization of the yield function \eqref{eq:yield_mcc} leads to:
\begin{equation}
\dfrac{2q}{M^2} \dfrac{\partial q}{\partial \Delta \boldsymbol{\varepsilon}} + \dfrac{\partial p}{\partial \Delta \boldsymbol{\varepsilon}} \left( p - p_c \right) + p \left( \dfrac{\partial p}{\partial \Delta \boldsymbol{\varepsilon}} - \dfrac{\partial p_c}{\partial \Delta \boldsymbol{\varepsilon}} \right) = 0 .
\label{eq:d_yield_mcc}
\end{equation}
\gm{Eq. \ref{eq:d_yield_mcc} 
allows to incorporate \eqref{eq:mcc_p_de_and_s_de} and \eqref{eq:mcc_pc_de_and_q_de} to \eqref{eq:d_yield_mcc}, which then effectively leads to an equation to determine $\dfrac{\partial \Delta \phi}{\partial \Delta \boldsymbol{\varepsilon}}$ for the MCC model:}
\begin{equation}
\left[ \dfrac{3}{M^2} \mathbf{s} : \mathbf{D} + \left( 2p - p_c \right) a - p b \right] \dfrac{\partial \Delta \phi}{\partial \Delta \boldsymbol{\varepsilon}} = -\left( 2p - p_c \right) \mathbf{A} + p \mathbf{B} - \dfrac{3}{M^2} \mathbf{s} : \mathbb{D} .
\label{eq:dphi_de_mcc}
\end{equation}

For the CASM model, linearization of the yield function \eqref{eq:yield_casm} leads to:
\begin{equation}
\dfrac{1}{p} \left[ \dfrac{1}{\ln R} - N \left( \dfrac{q}{Mp} \right)^N \right] \dfrac{\partial p}{\partial \Delta \bs{\varepsilon}} + \dfrac{N q^{N-1}}{(M p)^N} \dfrac{\partial q}{\partial \Delta \bs{\varepsilon}} - \dfrac{1}{\ln R} \dfrac{1}{p_c} \dfrac{\partial p_c}{\partial \Delta \bs{\varepsilon}} = 0 ,
\label{eq:d_yield_casm}
\end{equation}

Substituting \eqref{eq:mcc_p_de_and_s_de} and \eqref{eq:mcc_pc_de_and_q_de} into \eqref{eq:d_yield_casm} leads to an equation to determine $\dfrac{\partial \Delta \phi}{\partial \Delta \boldsymbol{\varepsilon}}$ for the CASM model:
\begin{equation}
\left( \dfrac{a}{p} \left[ \dfrac{1}{\ln R} - N \left( \dfrac{q}{Mp} \right)^N \right] + \dfrac{3}{2} \dfrac{N q^{N-2} (\mathbf{s} : \mathbf{D})}{(M p)^N} - \dfrac{1}{\ln R} \dfrac{b}{p_c} \right) \dfrac{\partial \Delta \phi}{\partial \Delta \boldsymbol{\varepsilon}} = - \dfrac{\mathbf{A}}{p} \left[ \dfrac{1}{\ln R} + N \left( \dfrac{q}{Mp} \right)^N \right] + \dfrac{3}{2} \dfrac{N q^{N-2} (\mathbf{s} : \mathbb{D})}{(M p)^N} + \dfrac{1}{\ln R} \dfrac{\mathbf{B}}{p_c} .
\label{eq:dphi_de_casm}
\end{equation}
Solving \eqref{eq:dphi_de_mcc} and \eqref{eq:dphi_de_casm}, respectively, the consistent tangent operator for both models can be calculated using \eqref{eq:mcc_p_de_and_s_de} and \eqref{eq:tangent_operator}. Details of the  derivation of the terms $\mathbf{A}, a, \mathbf{B}, b, \mathbf{D}$ and $\mathbb{D}$, are contained in the  \Cref{app:mcc_tangent} for the MCC model, and in the \Cref{app:casm_tangent} for the CASM model.

\subsection{Sub-stepping integration scheme}
\label{sec:mcc_multi}
\subsubsection{Stress integration algorithm}
%
%\todo{make a figure for multi-step (line graph with step marking)}

%\todo{rephrase this paragraph} We assume the case that the local load step is divided by two equally in size, which means the strain increment $\Delta \boldsymbol{\varepsilon} = 2 \Delta \tilde{\boldsymbol{\varepsilon}}$. We denotes ${}^k(*)$ is the value at the end of the $k^{\text{th}}$ sub-step. In that sense, the final value $(*)={}^2(*)$ for a two-step scheme.
%
In the sub-stepping integration scheme, the strain increment $\Delta \boldsymbol{\varepsilon}$ is divided into sub-increments
\begin{equation}
\Delta \boldsymbol{\varepsilon} = \sum_{k=1}^m {}^{k} \Delta \boldsymbol{\varepsilon} \, ,
\qquad
{}^{k} \Delta \boldsymbol{\varepsilon} = \alpha_k \Delta \boldsymbol{\varepsilon} \, ,
\qquad
\sum_{k=1}^m \alpha_k = 1 .
\label{eq:sub_step_rule}
\end{equation}
where  $m$ is the number of sub-steps. 
% to be differentiated with $n$, which denotes the previous time step.
%
The stress integration in one sub-step is the same as in the one-step scheme in \Cref{sec:mcc_one}, with the initial state of sub-step $k+1$ defined as the state of the previous sub-step $k$.  The stress at the end of the sub-step $k$ is denoted as 
${^k} \boldsymbol{\sigma}$. 
The final stress is the stress obtained in sub-step $k=m$: $\bs{\sigma}_{n+1} = {^m} \bs{\sigma}_{n+1}$.
Evidently, the one-step algorithm is a particular case of the sub-stepping algorithm, with the number of sub-steps set to  one.

\subsubsection{Consistent tangent operator}
The tangent operator $\dfrac{ \partial \, {^k} \boldsymbol{\sigma}}{\partial \, {^k} \Delta \boldsymbol{\varepsilon}}$ can be computed using the approach of one-step scheme. Nevertheless, the tangent operator $\dfrac{\partial \boldsymbol{\sigma}}{\partial \Delta \boldsymbol{\varepsilon}}=\dfrac{\partial \, {^m} \boldsymbol{\sigma}}{\partial \Delta \boldsymbol{\varepsilon}}$ canot  be trivially computed as the sum of the tangent operators in the individual sub-step, since ${^{k+1}} \boldsymbol{\sigma}$ depends on ${^k} \boldsymbol{\sigma}$ and ${^{k+1}} \Delta \boldsymbol {\varepsilon}$. Instead, $\dfrac{\partial \, {^k} \boldsymbol{\sigma}}{\partial \Delta \boldsymbol{\varepsilon}}$ shall be determined for all sub-steps. 

From \eqref{eq:tangent_operator}, one obtains  $\dfrac{\partial \, {^k} \boldsymbol{\sigma}}{\partial \Delta \boldsymbol{\varepsilon}} = \dfrac{\partial \, {^k} \mathbf{s}}{\partial \Delta \boldsymbol{\varepsilon}} + \mathbf{I} \otimes \dfrac{\partial \, {^k} p}{\partial \Delta \boldsymbol{\varepsilon}}$.  $\dfrac{\partial \, {^k} \mathbf{s}}{\partial \Delta \boldsymbol{\varepsilon}}$ and $\dfrac{\partial \, {^k} p}{\partial \Delta \boldsymbol{\varepsilon}}$ can be determined recursively, as presented in the following:
Firstly, $\dfrac{\partial \, {}^k \mathbf{s}}{\partial \, {}^k \Delta \boldsymbol{\varepsilon}}$, $\dfrac{\partial \, {}^k p}{\partial \, {}^k \Delta \boldsymbol{\varepsilon}}$, $\dfrac{\partial \, {}^k p_c}{\partial \, {}^k \Delta \boldsymbol{\varepsilon}}$ with $k=1$ can be determined using the one-step scheme in \Cref{sec:mcc_one}. Applying \eqref{eq:sub_step_rule} leads to the consistent linearization of the stresses in sub-step $k=1$:
$\dfrac{\partial \, {}^1 \mathbf{s}}{\partial \Delta \boldsymbol{\varepsilon}} = \alpha_1 \dfrac{\partial \, {}^1 \mathbf{s}}{\partial \, {}^1 \Delta \boldsymbol{\varepsilon}}$, $\dfrac{\partial \, {}^1 p}{\partial \Delta \boldsymbol{\varepsilon}} = \alpha_1 \dfrac{\partial \, {}^1 p}{\partial \,  {}^1 \Delta \boldsymbol{\varepsilon}}$ and $\dfrac{\partial \, {}^1 p_c}{\partial \Delta \boldsymbol{\varepsilon}} = \alpha_1 \dfrac{\partial \, {}^1 p_c}{\partial \, {}^1 \Delta \boldsymbol{\varepsilon}}$.
Secondly, assuming $\dfrac{\partial \, {}^k \mathbf{s}}{\partial \Delta \boldsymbol{\varepsilon}}$, $\dfrac{\partial \, {}^k p}{\partial \Delta \boldsymbol{\varepsilon}}$ and $\dfrac{\partial \, {}^k p_c}{\partial \Delta \boldsymbol{\varepsilon}}$ are already known from the previous sub-step, the linearization in the current sub-step $k+1$ is formulated depending, if the current sub-step is an elastic or an elasto-plastic step: 

\underline{\textit{Elastic state:}}

If the sub-step $(k+1)$ is elastic, the following relations hold:
\begin{equation}
{}^{k+1} p = {}^{k} p \, \text{exp} \left( \dfrac{1+e}{\kappa} {}^{k+1} \Delta \epsilon_v \right) \, ,
\quad
{}^{k+1} \mathbf{s} = {}^k \mathbf{s} + 2 \bar{\mu} \, {}^{k+1} \Delta \bs{\varepsilon}_d \, ,
\quad
\bar{\mu} = \dfrac{{}^{k+1} p - {}^k p}{{}^{k+1} \Delta \epsilon_v} r .
\label{eq:mcc_elastic_stresses_s}
\end{equation}

Taking the linearization of \eqref{eq:mcc_elastic_stresses_s} leads to
\begin{equation}
\dfrac{\partial \, {}^{k+1} p}{\partial \Delta \boldsymbol{\varepsilon}} = \dfrac{{}^{k+1} p}{{}^{k} p} \dfrac{\partial {}^k p}{\partial \Delta \bs{\varepsilon}} + \alpha_{k+1} \, {}^{k+1} K \mathbf{I} \, ,
\quad
\dfrac{\partial \, {}^{k+1} \mathbf{s}}{\partial \Delta \boldsymbol{\varepsilon}} = \dfrac{\partial \, {}^{k} \mathbf{s}}{\partial \Delta \boldsymbol{\varepsilon}} + 2 \alpha_{k+1} \bar{\mu} \mathbb{I}_d + 2 \, {}^{k+1} \Delta \bs{\varepsilon}_d \otimes \dfrac{\partial \bar{\mu}}{\partial \Delta \bs{\varepsilon}},
\end{equation}
with
\begin{equation}
\dfrac{\partial \bar{\mu}}{\partial \Delta \bs{\varepsilon}} = \dfrac{\bar{\mu}}{{}^k p} \dfrac{\partial \, {}^k p}{\partial \Delta \bs{\varepsilon}} + \alpha_{k+1} \dfrac{{}^{k+1} K r - \bar{\mu}}{{}^{k+1} \Delta \epsilon_v} \mathbf{I} \, ,
\qquad
{}^{k+1} K = \dfrac{1+e}{\kappa} {}^{k+1} p .
\end{equation}

\underline{\textit{Elasto-plastic state:}}
The linearizations $\dfrac{\partial \, {}^{k+1} \mathbf{s}}{\partial \Delta \boldsymbol{\varepsilon}}$, $\dfrac{\partial \, {}^{k+1} p}{\partial \Delta \boldsymbol{\varepsilon}}$ and $\dfrac{\partial \, {}^{k+1} p_c}{\partial \Delta \boldsymbol{\varepsilon}}$ in the plastic state are re-written in terms of $\dfrac{\partial \, {}^{k+1} \Delta \phi}{\partial \Delta \boldsymbol{\varepsilon}}$.
To simplifiy notation, $\dfrac{\partial \, {}^{k+1} \Delta \phi}{\partial \Delta \boldsymbol{\varepsilon}}$ will be denoted as $\dfrac{\partial \Delta \phi}{\partial \Delta \boldsymbol{\varepsilon}}$ for brevity:
\begin{equation}
\dfrac{\partial \, {}^{k+1} p}{\partial \Delta \boldsymbol{\varepsilon}} = \mathbf{A} + a \dfrac{\partial \Delta \phi}{\partial \Delta \boldsymbol{\varepsilon}} \, ,
\qquad
\dfrac{\partial \, {}^{k+1} p_c}{\partial \Delta \boldsymbol{\varepsilon}} = \mathbf{B} + b \dfrac{\partial \Delta \phi}{\partial \Delta \boldsymbol{\varepsilon}} \, ,
\qquad
\dfrac{\partial \, {}^{k+1} \mathbf{s}}{\partial \Delta \boldsymbol{\varepsilon}} = \mathbb{D} + \mathbf{D} \otimes \dfrac{\Delta \phi}{\Delta \boldsymbol{\varepsilon}} .
\end{equation}
The procedure to compute $\dfrac{\partial \Delta \phi}{\partial \Delta \boldsymbol{\varepsilon}}$ is analogous to the one-step scheme for the elasto-plastic state (see \Cref{sec:mcc_tangent}). The derivation of the terms $\mathbf{A}, a, \mathbf{B}, b, \mathbf{D}$ and $\mathbb{D}$ for the sub-step $k+1$ can be found in \Cref{app:mcc_tangent_substepping} for the MCC model, and \Cref{app:casm_tangent_substepping} for the CASM model.

%\gm{*** GM *** BIS HIERHER ***}

\section{Numerical examples}
\label{sec:numerical_examples}

\subsection{Triaxial test of a soil sample}

%In the first example, the drained compressive triaxial test on a cube sample is performed to validate the accuracy of the implicit sub-stepping algorithm. The geometry and boundary condition can be seen in \Cref{fig:triaxial}, in which the top surface is applied with distributed pressure $p_v$ and the sides are applied with pressure $p_h$ for load control procedure and respectively $u_v$, $u_h$ if the displacement control procedure is used. The geometry is meshed with single hexahedra element with 8 nodes. The test is performed on 3 clay samples with the initial stress state as normally consolidated ($\text{OCR}=1$), lightly over-consolidated ($\text{OCR}=2$) and highly over-consolidated ($\text{OCR}=5$). The proportional loading scheme is selected as the controlled stress path, for which the analytical solution exists. For MCC model, the analytical solution subjected to drained proportional loading can be found in \citep{Peric:06}, and for the CASM model, in \Cref{app:casm_load_sol}. In the proportional loading scheme, the load is controlled in such a way that $\dot{q} = k \dot{p}$ ($k$: loading rate). If the displacement control scheme is used, the axial strain $\epsilon_a$ is controlled to match with the displacement history of the analytical solution.

\begin{figure}[htb!]
\centering
\begin{minipage}{0.5\textwidth}
  \includegraphics[trim=0.5cm 0cm 0cm 0cm,clip,scale=1.0]{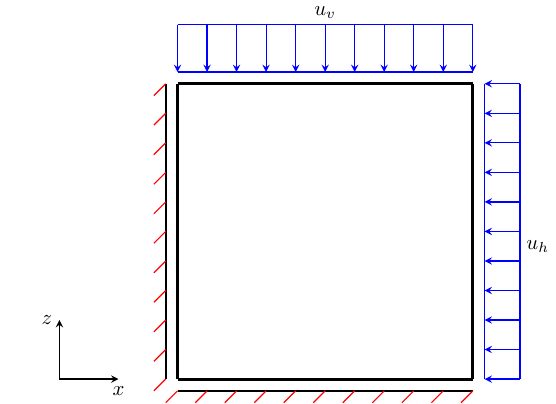}
\end{minipage}%
\begin{minipage}{0.5\textwidth}
  \includegraphics[trim=0.5cm 0cm 0cm 0cm,clip,scale=1.0]{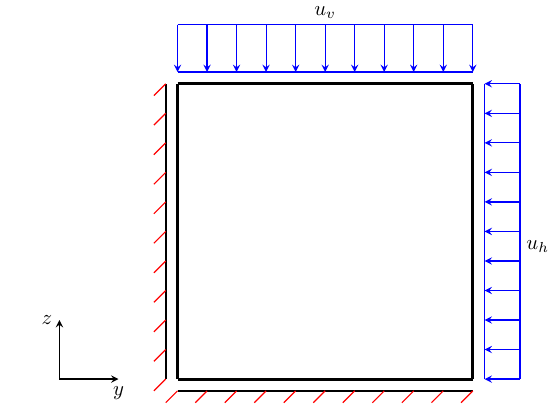}
\end{minipage}
\caption{Triaxial test example: Illustration of geometry and boundary conditions.}
\label{fig:triaxial}
\end{figure}

In the first example, the drained compressive triaxial test on a cube sample is performed to validate the accuracy of the implicit sub-stepping algorithm. The geometry and boundary condition can be seen in \Cref{fig:triaxial}, in which the top surface is applied with prescribed displacement $u_v$ and the sides are applied with prescribed displacement $u_h$. The geometry is meshed with single hexahedra element with 8 nodes. The test is performed on 3 clay samples with the initial stress state as normally consolidated (case 1, $\text{OCR}=1$), lightly over-consolidated (case 2, $\text{OCR}=2$) and highly over-consolidated (case 3, $\text{OCR}=5$). The proportional loading scheme is selected as the controlled stress path, for which the analytical solution exists. For MCC model, the analytical solution subjected to drained proportional loading can be found in \citep{Peric:06}, and for the CASM model, in \Cref{app:casm_load_sol}. In the proportional loading scheme, the load is controlled in such a way that $\dot{q} = k \dot{p}$ ($k$: loading rate). To produce the stress path that matches with the analytical solution, the axial strain $\epsilon_a$ is controlled to match with the displacement history of the analytical solution. It is noted that $\epsilon_a=\dfrac{1}{3} \epsilon_v + \epsilon_q$, in which $\epsilon_v$ is the volumetric strain and $\epsilon_q$ is the shear strain.

In the simulation, the following parameters are used for the MCC model: \{ $\lambda=0.066$, $\kappa=0.0077$, $M=1.2$, $e=1.788$, $\nu=0.3$ \}; and for the CASM model: \{ $\lambda=0.066$, $\kappa=0.0077$, $N=3$, $M=1.2$, $R=2.0$, $e=1.788$, $\nu=0.3$ \}. The implicit sub-stepping algorithm is enforced to run with 2 sub-steps. The loading rate is selected as $k=3$.

\subsubsection{Results with MCC model}

\begin{figure}[htb!]
\centering
\begin{minipage}{0.55\textwidth}
  \includegraphics[trim=0.0cm 0cm 0cm 0cm,clip,scale=1.0]{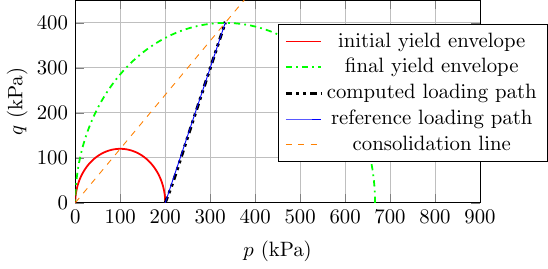}
\end{minipage}%
\begin{minipage}{0.45\textwidth}
  \includegraphics[trim=0.0cm 0cm 0cm 0cm,clip,scale=1.0]{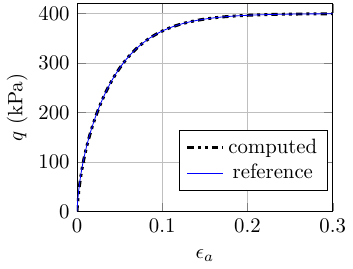}
\end{minipage}
\caption{Triaxial test example, MCC model, case 1: Left) computed stress path and Right) Load-displacement curve.}
\label{fig:mcc_example_1}
\end{figure}

\begin{figure}[htb!]
\centering
\begin{minipage}{0.55\textwidth}
  \includegraphics[trim=0.0cm 0cm 0cm 0cm,clip,scale=1.0]{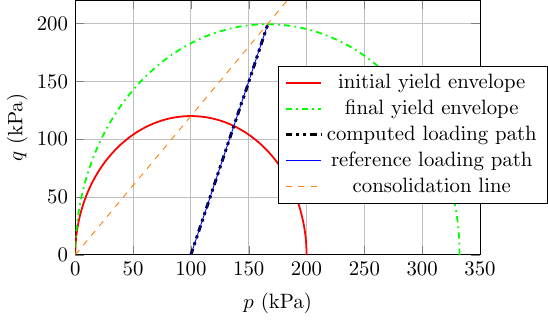}
\end{minipage}%
\begin{minipage}{0.45\textwidth}
  \includegraphics[trim=0.0cm 0cm 0cm 0cm,clip,scale=1.0]{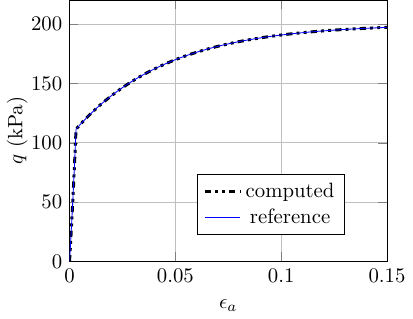}
\end{minipage}
\caption{Triaxial test example, MCC model, case 2: Left) computed stress path and Right) Load-displacement curve.}
\label{fig:mcc_example_2}
\end{figure}

\begin{figure}[htb!]
\centering
\begin{minipage}{0.55\textwidth}
  \includegraphics[trim=0.0cm 0cm 0cm 0cm,clip,scale=1.0]{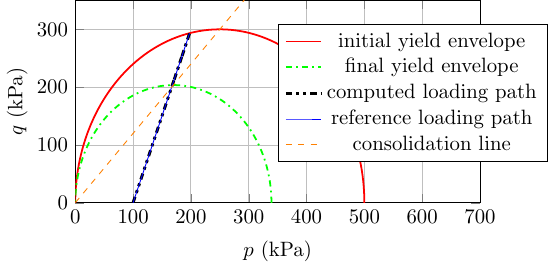}
\end{minipage}%
\begin{minipage}{0.45\textwidth}
  \includegraphics[trim=0.0cm 0cm 0cm 0cm,clip,scale=1.0]{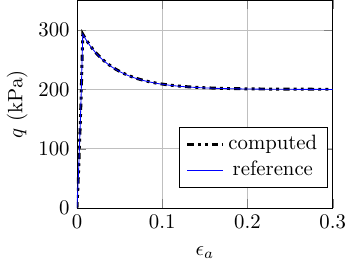}
\end{minipage}
\caption{Triaxial test example, MCC model, case 3: Left) computed stress path and Right) Load-displacement curve.}
\label{fig:mcc_example_3}
\end{figure}

The results, as presented in \Cref{fig:mcc_example_1} for normally consolidated soil, \Cref{fig:mcc_example_2} for lightly consolidated soil and \Cref{fig:mcc_example_3} for highly consolidated soil show excellent agreement with the analytical results. In case 1, the stress state starts from the initial yield envelope and stops at the consolidation line as expected. The shear stress asymptotically converges and the soil sample cannot sustain further load. \Cref{fig:mcc_example_1} shows the computed stress path (Left) and the according load-displacement curve (Right). The same behaviour is observed for case 2, where the stress state starts from the elastic domain and enters the plastic state after hitting the initial yield envelope (see \Cref{fig:mcc_example_2} (Left)). From that point, the soil sample starts to harden and the shear stress asymptotically converges (see \Cref{fig:mcc_example_2} (Right)). Case 1 and case 2 show that the implicit sub-stepping algorithm works well on the dry side of the modified Cam-Clay model.

In case 3, the softening behaviour on the wet side of the MCC is tested. As can be seen in \Cref{fig:mcc_example_3} (Left), the stress state starts in the elastic domain and after hitting the initial yield envelope, it starts to soften and the shear stress asymptotically converges (see \Cref{fig:mcc_example_3} (Right)). The stress state ultimately stops at the consolidation line as expected.

\subsubsection{Results with CASM model}

\begin{figure}[htb!]
\centering
\begin{minipage}{0.55\textwidth}
  \includegraphics[trim=0.0cm 0cm 0cm 0cm,clip,scale=1.0]{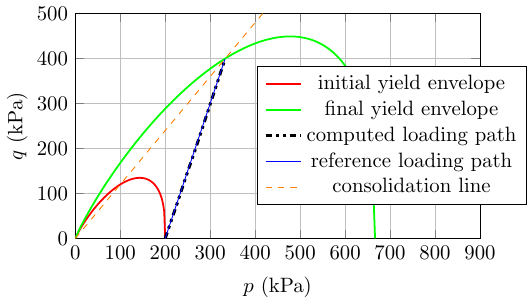}
\end{minipage}%
\begin{minipage}{0.45\textwidth}
  \includegraphics[trim=0.0cm 0cm 0cm 0cm,clip,scale=1.0]{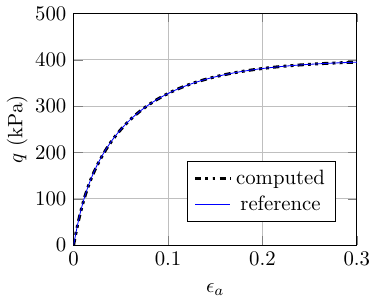}
\end{minipage}
\caption{Triaxial test example, CASM model, case 1: Left) computed stress path and Right) Load-displacement curve.}
\label{fig:casm_example_1}
\end{figure}

\begin{figure}[htb!]
\centering
\begin{minipage}{0.55\textwidth}
  \includegraphics[trim=0.0cm 0cm 0cm 0cm,clip,scale=1.0]{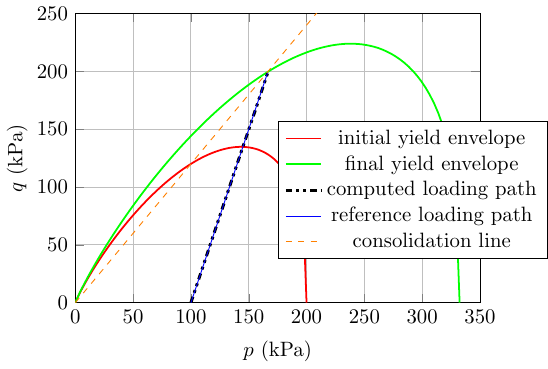}
\end{minipage}%
\begin{minipage}{0.45\textwidth}
  \includegraphics[trim=0.0cm 0cm 0cm 0cm,clip,scale=1.0]{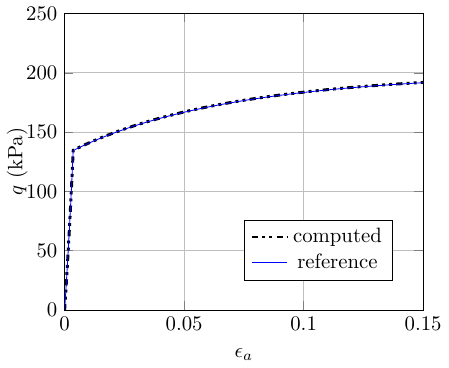}
\end{minipage}
\caption{Triaxial test example, CASM model, case 2: Left) computed stress path and Right) Load-displacement curve.}
\label{fig:casm_example_2}
\end{figure}

\begin{figure}[htb!]
\centering
\begin{minipage}{0.55\textwidth}
  \includegraphics[trim=0.0cm 0cm 0cm 0cm,clip,scale=1.0]{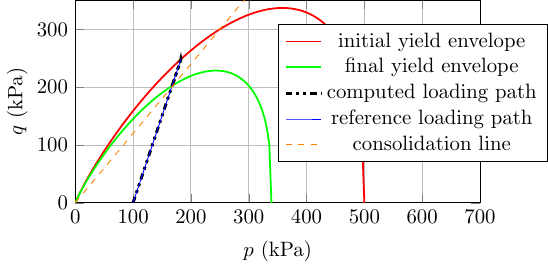}
\end{minipage}%
\begin{minipage}{0.45\textwidth}
  \includegraphics[trim=0.0cm 0cm 0cm 0cm,clip,scale=1.0]{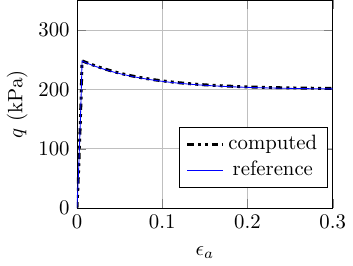}
\end{minipage}
\caption{Triaxial test example, CASM model, case 3: Left) computed stress path and Right) Load-displacement curve.}
\label{fig:casm_example_3}
\end{figure}

The implicit sub-stepping algorithm also work expectedly with the CASM model in the triaxial test, as shown in \Cref{fig:casm_example_1} (case 1), \Cref{fig:casm_example_2} (case 2) and \Cref{fig:casm_example_3} (case 3). For each case, the evolution of the stress state is analogous to the respective case of MCC model. Nevertheless, the stress state hits the initial yield envelope early for case 3 and lately in case 2 due to the shape of the yield surface. Similar to the MCC case, the implicit sub-stepping algorithm also performs well on the wet side of the CASM model.

\subsection{Excavation of a soil block}

In this example, the simulation of the excavation process of a soil block is presented to demonstrate the use of the implicit sub-stepping integration algorithm of MCC and CASM model in a practical application. The geometry and material parameters for this example are taken from \citep{Borja:91a}. For MCC model, the material parameters read: $\lambda=0.37$, $\kappa=0.054$, $M=1.4$, $e=2.52$ and $\nu=0.35$. For CASM model, additional parameters are set, in which $N=1.4$ and $R=2.0$, to approximately match the corresponding MCC model. In difference to the original example, the over-consolidation ratio ($\text{OCR}$) is preset to 1.7 for the half upper part and $\text{OCR}=1.0$ for the half lower part of the soil block. The initial stress of the soil are generated by $K_0$-procedure with $K_0=1.2$ and the soil density is $\rho=1732 \, kg/m^3$. The preconsolidation pressure are generated from the initial stress using the yield criteria and $\text{OCR}$ as below:
\begin{itemize}
\item for MCC:
\begin{equation}
p_c = \text{OCR} \left( p + \dfrac{q^2}{M^2 p} \right) .
\label{eq:pc_ocr_mcc}
\end{equation}
\item for CASM:
\begin{equation}
p_c = \text{OCR} \, p \, \exp \left[ \ln R \left( \dfrac{q}{M p} \right)^N \right] .
\label{eq:pc_ocr_casm}
\end{equation}
\end{itemize}

The sub-stepping algorithm is used by default, and the number of sub-steps are fixed to 4. Using the same analysis sequence as the original example\citep{Borja:91a}, the analysis with different number of steps are simulated. With a total of 8 element layers representing the excavation domain, the number of excavation steps (lifts) are simulated as 8, 4, 2, 1 corresponding to 1, 2, 4, 8 number of excavated layers per excavation step, respectively. The displacements of reference point A \citep{Borja:91a} corresponding to different lifts, together with the number of average iterations in an excavation step, are reported in \Cref{tab:block_excavation}. It is noted that, the displacements are presented in \textit{mm} and the tolerance for convergence is $10^{-13}$.

\begin{table}[h!]
\centering
\begin{tabular}{cccccccc}
& \multicolumn{3}{c}{MCC} & & \multicolumn{3}{c}{CASM} \\
\cline{2-4}  \cline{6-8} 
No. of lifts & $d_u$ & $d_v$ & No. iter/lift & & $d_u$ & $d_v$ & No. iter/lift \\
\hline
1 & -29.45 & -16.76 & 9 & & -30.28 & -15.53 & 9 \\
2 & -28.67 & -16.47 & 7 & & -29.58 & -15.38 & 7 \\
4 & -28.18 & -16.37 & 6 & & -29.09 & -15.29 & 6.2 \\
8 & -28.02 & -16.25 & 5.5 & & -28.85 & -15.11 & 5.6 \\
\hline
\end{tabular}
\caption{Excavation of a soil block: Displacements at the reference point A \citep{Borja:91a} and the convergence of the implicit sub-stepping integration algorithm for different number of lifts.}
\label{tab:block_excavation}
\end{table}

From \Cref{tab:block_excavation}, one can see that the difference in displacements measured at the reference point of MCC and CASM model are not large. This is expected since the yield surface of MCC and CASM are approximately matched. In addition, the convergence of the implicit sub-stepping algorithm exhibits quadratic convergence behaviour, which take a minimum of 5 iteration and maximum 9 iterations to converge in a single load step.

%\subsection{Tunnel excavation example}
%
%\begin{figure}[htb!]
%\centering
%\includegraphics[trim=0.0cm 0cm 0cm 0cm,clip,scale=0.7]{figs/whl_2d.pdf}
%\caption{Tunnel excavation example: Illustration of geometry and boundary conditions.}
%\label{fig:whl_2d}
%\end{figure}
%
%In this section, a synthetic tunnel example
%
%\begin{figure}[htb!]
%\centering
%\includegraphics[trim=0.0cm 0cm 0cm 0cm,clip,scale=1.0]{fig_results/whl/whl_2d_settlement.pdf}
%\caption{Tunnel excavation example: Settlement computed with different critical state models.}
%\label{fig:whl_2d_settlement}
%\end{figure}

\subsection{Tunnel excavation example}

In the third example, a simulation of tunnel excavation employing NATM method is presented. The objective of this example is to illustrate the advantage of implicit sub-stepping algorithm over the one-step algorithm in which the latter one fails to work, which means the implicit sub-stepping algorithm does not converge in one step. The criteria of the local failure is either the non-convergence of the local Newton-Raphson iteration to solve the local nonlinear system, or the inadmissibility of the resulting stress state, i.e. $p \leq 0 \lor q < 0 \lor p_c \leq 0$.

\begin{figure}[htb!]
\centering
\includegraphics[trim=0.0cm 0cm 0cm 0cm,clip,scale=1.0]{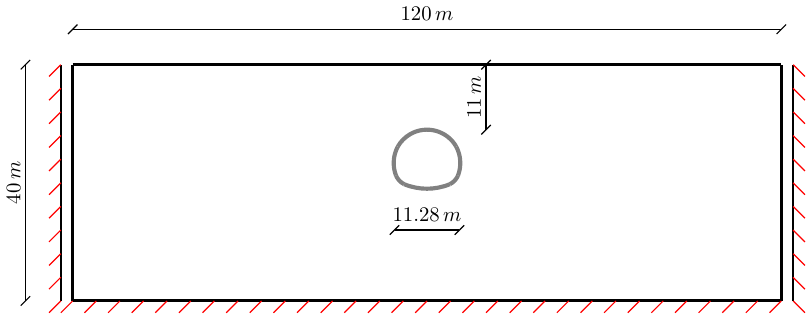}
\caption{Tunnel excavation example: Illustration of geometry and boundary conditions.}
\label{fig:natm_2d}
\end{figure}

\Cref{fig:natm_2d} shows the geometry and boundary conditions of the problem at hand. They are adopted from the tutorial 7 of ADONIS geotechnical software \citep{Mikola:17}. The following parameters are adopted for the MCC and CASM model: $\lambda=0.147$, $\kappa=0.06$, $M=1.05$, $e=3.56$ and $\nu=0.3$. For CASM, additional parameters are used: $N=1.4$ and $R=2.0$ to approximate the respective MCC model. The geometry of the problem is meshed by 5231 quadratic quadrilateral element (9-node quad) and the full integration scheme with 9 integration points per element is used.

Because the number of required sub-steps is not known in advance, an adaptive stepping scheme is used in accordance with the implicit sub-stepping algorithm. If the integration at a sub-step fails, the number of sub-steps at the integration point will be doubled and the sub-stepping integration is started over.
%It is noted that the adaptive sub-stepping is allowed only locally and does not alter the global iteration. In addition, an occurrence of more than one sub-steps in the adaptive stepping scheme is the indicator that the one-step algorithm will fail, because the one-step algorithm is a particular case of sub-stepping algorithm, as described in \Cref{sec:mcc_multi}.
Tolerance of $10^{-8}$ is adopted for the local iteration and tolerance of $10^{-10}$ is adopted for the global Newton-Raphson loop.

Analogous to the previous example, the initial stress state is generated using $K_0$-procedure with $K_0=0.5$, $\rho=1732 \, kg/m^3$ and the initial preconsolidation pressure profile is computed using \Cref{eq:pc_ocr_mcc} for MCC and \Cref{eq:pc_ocr_casm} for CASM respectively. The over-consolidation ratio is set to $\text{OCR}=2.0$.

%To simulate the excavation process, a stress relaxation approach is adopted, in which the traction is applied on the tunnel wall after the elements in the excavation domain are deactivated. The traction on the tunnel wall is computed from the initial internal stress of the soil and is reduced to 40 \% when the shotcrete is installed in the current configuration. The shotcrete is modelled as beam element with $E=1.56e6 \, kPa$, $\nu=0.2$ and the thickness of the beam is $t=0.1 \, m$. After the shotcrete is installed, the applied traction on tunnel wall is removed. The model is assumed to be in plane strain condition. The settlement profile on the top surface, after the shotcrete is installed and applied traction is vanished, is shown in \Cref{fig:natm_2d_settlement}. One can see that, the results of MCC and CASM is very closed, due to the yield surfaces are matched approximately.

To simulate the excavation process, a stress relaxation approach is adopted. At first, the elements in the excavation domain are deactivated and the corresponding traction is applied on the tunnel wall to maintain the equilibrium. The traction is computed from the initial stress field within the soil. In the second step, the traction is reduced to 40 \%. Subsequently, the shotcrete, represented by beam elements is installed in the current configuration. The parameters for the shotcrete beam element are $E=1.56e6 \, kPa$, $\nu=0.2$ and the thickness is $t=0.1 \, m$. Finally, after the shotcrete is installed, the applied traction on tunnel wall is removed. The  plane strain condition is assumed in the analysis.

\begin{figure}[htb!]
\centering
\includegraphics[trim=0.0cm 0cm 0cm 0cm,clip,scale=1.0]{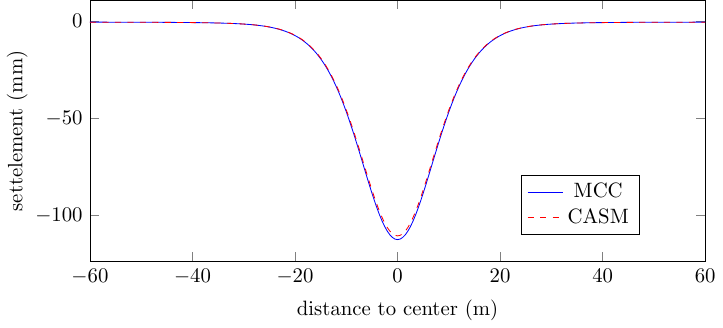}
\caption{Tunnel excavation example: Settlement profile on top surface.}
\label{fig:natm_2d_settlement}
\end{figure}

The settlement profile on the top surface, after the final step, is shown in \Cref{fig:natm_2d_settlement}. As one shall expect, the results of MCC and CASM is very closed, due to the yield surfaces are matched.

%In terms of convergence, the single step fails when the traction is 40 \% of the initial value for both MCC and CASM, whereby for the implicit sub-stepping algorithm, a total of 70 sub-steps for MCC and 392 sub-steps for CASM is required at failed plastic point. The statistics of the sub-stepping algorithm are summarized in \Cref{tab:natm_stat}. Here it is denoted that PP is plastic point, and SPP as sub-stepping plastic point, by which the one-step algorithm fails and the adaptive sub-stepping scheme is applied. The average number of sub-steps is computed by summing all number of sub-steps at SPP and divide by the number of SPPs.

In terms of convergence, the classical one-step algorithm fails for both MCC and CASM in the loading step when the traction is reduced. In the implicit sub-stepping algorithm, a total of 70 sub-steps for MCC and 392 sub-steps for CASM is required at those failed plastic points, recorded in the one-step analysis. The statistics of the sub-stepping algorithm are summarized in \Cref{tab:natm_stat}. In the table, PP denotes the plastic point, and SPP is the plastic point where the sub-stepping is required. The average number of sub-steps is computed by summing all number of sub-steps at SPP and divide by the number of SPPs.

\begin{table}[h!]
\centering
\begin{tabular}{cccccccccccc}
& \multicolumn{5}{c}{MCC} & & \multicolumn{5}{c}{CASM} \\
\cline{2-6}  \cline{8-12}
& & & \multicolumn{3}{c}{No. of sub-steps} & & & & \multicolumn{3}{c}{No. of sub-steps} \\
\cline{4-6}  \cline{10-12}
Traction & No. of PP & No. of SPP & Min. & Avg. & Max. & & No. of PP & No. of SPP & Min. & Avg. & Max \\
\hline
70 \% & 45450 & 0 & 0 & 0 & 0 & & 45450 & 0 & 0 & 0 & 0 \\
40 \% & 45450 & 13 & 2 & 5.4 & 32 & & 45450 & 87 & 2 & 4.5 & 16 \\
26.67 \% & 45450 & 0 & 0 & 0 & 0 & & 45450 & 0 & 0 & 0 & 0 \\
13.33 \% & 45450 & 0 & 0 & 0 & 0 & & 45450 & 0 & 0 & 0 & 0 \\
0 \% & 45450 & 0 & 0 & 0 & 0 & & 45450 & 0 & 0 & 0 & 0 \\
\hline
\end{tabular}
\caption{Excavation of a soil block: Displacements at the reference point A and the convergence of the implicit sub-stepping integration algorithm for different number of lifts.}
\label{tab:natm_stat}
\end{table}

%The results show that the critical phase of loading is when traction is 40 \% of initial value, where the sub-stepping integration is required to solve the failed plastic points. After the shotcrete is installed, the stress from the soil is carried by the beam and no further sub-stepping is required. Although the number of SPP is not high, it is crucial for the continuation of the loading and ultimately the simulation can get pass the tricky situation where one-step integration cannot handle.

In this example, the sub-stepping integration is required only in the step where the reduced traction occurs. Although the number of SPP is not high, it is crucial for the continuation of the analysis where one-step algorithm fails and cannot continue.

\section{Conclusions}
\label{sec:concl}

This paper presents an implicit sub-stepping algorithm that greatly improves the computability of the critical state soil models. The sub-stepping stress integration is not new, nevertheless this is the first time a sub-stepping algorithm is fully linearized to be used with the implicit Newton-Raphson iteration for the critical state models, particularly for the popular Modified Cam-Clay and Clay-And-Sand models. The validation examples show that the sub-stepping integration algorithm can reproduce the analytical solution and the results exhibit quadratic convergence behaviour. In addition, the implicit sub-stepping algorithm is able to integrate the plastic points where the traditional one-step scheme fails. This greatly enhances the robustness of the critical state models and further promotes the usage of critical state models in geotechnical analysis. The implicit sub-stepping integration scheme can be extended to other soil models based on critical state concept, such as bounding Cam-Clay model, Barcelona-Basic model, etc, and will be the subject for future research.

\subsection{Acknowledgements} This presented work was conducted in the framework of the Collaborative Research Project SFB 837 "Interaction Modeling in Mechanized Tunneling", financed by the German Research Foundation (DFG) (Project number 77309832). The authors would like to thank the DFG for the support of this project.
%Financial support was provided by the German Science Foundation (DFG) in the framework of project C1 of the Collaborative Research Center SFB 837. This support is gratefully acknowledged.

%\bibliographystyle{wileyNJD-VANCOUVER}
\bibliography{sd-lit,lit-add}
%\bibliography{sd-lit_seafile}

%\subsection{Bibliography}
%
%\begin{enumerate}[1]
%\item Use \verb"\bibliography{wileyNJD-AMA}" BST file for AMA reference style
%\item Use \verb"\bibliography{wileyNJD-APA}" BST file for APA reference style
%\item Use \verb"\bibliography{wileyNJD-AMS}" BST file for AMS reference style
%\item Use \verb"\bibliography{wileyNJD-VANCOUVER}" BST file for Vancouver reference style
%\item Use \verb"\bibliography{wileyNJD-ACS}" BST file for Chemistry reference style
%\end{enumerate}
%
%The normal commands for producing the reference list are:
%
%\begin{verbatim}
%\begin{thebibliography}{99}
%\bibitem{<x-ref label>}
%         <Reference details>
%.
%.
%.
%\end{thebibliography}
%\end{verbatim}
%
%\subsection{Appendix Section}
%
%\begin{verbatim}
\appendix

\section{Linearization of the one-step scheme, modified Cam-Clay model}
\label{app:mcc_tangent}

Taking the linearization of \eqref{eq:p_mcc} and \eqref{eq:pc_mcc}, one obtains:
\begin{align}
\dfrac{\partial p}{\partial \Delta \boldsymbol{\varepsilon}} &= \dfrac{1+e}{\kappa} p \dfrac{\partial \Delta \varepsilon^e_v}{\partial \Delta \boldsymbol{\varepsilon}} , \label{eq:p_de_1}
\\
\dfrac{\partial p_c}{\partial \Delta \boldsymbol{\varepsilon}} &= \theta p_c \left[ \Delta \phi \left( 2\dfrac{\partial p}{\partial \Delta \boldsymbol{\varepsilon}} - \dfrac{\partial p_c}{\partial \Delta \boldsymbol{\varepsilon}} \right) + \left( 2p - p_c \right) \dfrac{\partial \Delta \phi}{\partial \Delta \boldsymbol{\varepsilon}} \right] . \label{eq:pc_de_1}
\end{align}

\Cref{eq:pc_de_1} is equivalent to:
\begin{equation}
\dfrac{\partial p_c}{\partial \Delta \boldsymbol{\varepsilon}} = \dfrac{\theta p_c}{1 + \theta p_c \Delta \phi} \left[ 2\Delta \phi \dfrac{\partial p}{\partial \Delta \boldsymbol{\varepsilon}} + \left( 2p - p_c \right) \dfrac{\partial \Delta \phi}{\partial \Delta \boldsymbol{\varepsilon}} \right] .
\label{eq:pc_de_2}
\end{equation}

It is noted that $\Delta \varepsilon^p_v = \Delta \phi (2p-p_c)$ and $\Delta \varepsilon^e_v = \Delta \epsilon_v - \Delta \phi (2p-p_c)$, hence:
\begin{equation}
\dfrac{\partial \Delta \varepsilon^e_v}{\partial \Delta \boldsymbol{\varepsilon}} = \mathbf{I} - \Delta \phi \left( 2 \dfrac{\partial p}{\partial \Delta \boldsymbol{\varepsilon}} - \dfrac{\partial p_c}{\partial \Delta \boldsymbol{\varepsilon}} \right) - \left( 2p - p_c \right) \dfrac{\partial \Delta \phi}{\partial \Delta \boldsymbol{\varepsilon}} .
\label{eq:de_e_v_de_1}
\end{equation}

Substituting \eqref{eq:de_e_v_de_1} and \eqref{eq:pc_de_2} into \eqref{eq:p_de_1} leads to:
\begin{equation}
\left( 1 + \dfrac{1+e}{\kappa} p \dfrac{2 \Delta \phi}{1 + \theta p_c \Delta \phi} \right) \dfrac{\partial p}{\partial \Delta \boldsymbol{\varepsilon}} = \dfrac{1+e}{\kappa} p \left[ \mathbf{I} - \dfrac{2p - p_c}{1 + \theta p_c \Delta \phi} \dfrac{\partial \Delta \phi}{\partial \Delta \boldsymbol{\varepsilon}} \right] .
\label{eq:p_de_2}
\end{equation}

\Cref{eq:p_de_2} can be rewritten as:
\begin{equation}
\dfrac{\partial p}{\partial \Delta \boldsymbol{\varepsilon}} = \mathbf{A}_1 + a_1 \dfrac{\partial \Delta \phi}{\partial \Delta \boldsymbol{\varepsilon}},
\quad
K = \dfrac{1+e}{\kappa} p,
\quad
a_2 = 1 + \left( \theta p_c + 2K \right) \Delta \phi,
\quad
\mathbf{A} = \mathbf{A}_1 = \dfrac{1 + \theta p_c \Delta \phi}{a_2} K \mathbf{I},
\quad
a = a_1 = -\dfrac{2p - p_c}{a_2} K .
\label{eq:p_de_3}
\end{equation}

Replaces \eqref{eq:p_de_3} into \eqref{eq:pc_de_2}, one obtains:
\begin{equation}
\dfrac{\partial p_c}{\partial \Delta \boldsymbol{\varepsilon}} = \mathbf{B}_1 + b_1 \dfrac{\partial \Delta \phi}{\partial \Delta \boldsymbol{\varepsilon}} ,
\qquad
\mathbf{B} = \mathbf{B}_1 = \dfrac{2\theta p_c \Delta \phi}{a_2} K \mathbf{I} ,
\qquad
b = b_1 = \dfrac{2p-p_c}{a_2} \theta p_c .
\label{eq:pc_de_3}
\end{equation}

The deviatoric pressure reads:
\begin{equation}
\mathbf{s} = \eta \left( \mathbf{s}_n + 2\bar{\mu} \Delta \boldsymbol{\varepsilon}_d \right) \, ,
\qquad
\bar{\mu} = \dfrac{p - p_n}{\Delta \epsilon_v^e} r \, ,
\qquad
\eta = \left( 1 + 6\bar{\mu}\dfrac{\Delta \phi}{M^2} \right)^{-1} .
\label{eq:s}
\end{equation}

Taking the linearization of \eqref{eq:s} with respect to $\Delta \boldsymbol{\varepsilon}$, one obtains:
\begin{equation}
\dfrac{\partial \mathbf{s}}{\partial \Delta \boldsymbol{\varepsilon}} = \left( \mathbf{s}_n + 2\bar{\mu}\Delta \boldsymbol{\varepsilon}_d \right) \otimes \dfrac{\partial \eta}{\partial \Delta \boldsymbol{\varepsilon}} + 2\eta \Delta \boldsymbol{\varepsilon}_d \otimes \dfrac{\partial \bar{\mu}}{\partial \Delta \boldsymbol{\varepsilon}} + 2\eta \bar{\mu} \dfrac{\partial \Delta \boldsymbol{\varepsilon}_d}{\partial \Delta \boldsymbol{\varepsilon}} .
\label{eq:s_de_1}
\end{equation}

In \eqref{eq:s_de_1}, this relation holds: $\dfrac{\partial \Delta \boldsymbol{\varepsilon}_d}{\partial \Delta \boldsymbol{\varepsilon}} = \mathbb{I}_d = \mathbb{I} - \dfrac{1}{3} \mathbf{I} \otimes \mathbf{I}$, where $\mathbb{I}$ is the fourth-order identity tensor. The linearization $\dfrac{\partial \bar{\mu}}{\partial \Delta \boldsymbol{\varepsilon}}$ and $\dfrac{\partial \eta}{\partial \Delta \boldsymbol{\varepsilon}}$ come as following:
\begin{equation}
\dfrac{\partial \bar{\mu}}{\partial \Delta \boldsymbol{\varepsilon}} = \mathbf{D}_1 + d_1 \dfrac{\partial \Delta \phi}{\partial \Delta \boldsymbol{\varepsilon}} \, ,
\qquad
\mathbf{D}_1 = \dfrac{K r - \bar{\mu}}{K \Delta \epsilon_v^e} \mathbf{A}_1 \, ,
\qquad
d_1 = \dfrac{K r - \bar{\mu}}{K \Delta \epsilon_v^e} a_1 ,
\label{eq:mu_bar_de}
\end{equation}
\begin{equation}
\dfrac{\partial \eta}{\partial \Delta \boldsymbol{\varepsilon}} = \mathbf{D}_2 + d_2 \dfrac{\partial \Delta \phi}{\partial \Delta \boldsymbol{\varepsilon}} \, ,
\qquad
\mathbf{D}_2 = -6 \eta^2 \dfrac{\Delta \phi}{M^2} \mathbf{D}_1 \, ,
\qquad
d_2 = -\dfrac{6 \eta^2}{M^2} \left( \Delta \phi d_1 + \bar{\mu} \right) .
\label{eq:eta_de}
\end{equation}

Substituting \eqref{eq:mu_bar_de} and \eqref{eq:eta_de} into \eqref{eq:s_de_1} results in $\dfrac{\partial \mathbf{s}}{\partial \Delta \boldsymbol{\varepsilon}}$:
\begin{equation}
\dfrac{\partial \mathbf{s}}{\partial \Delta \boldsymbol{\varepsilon}} = \mathbb{D} + \mathbf{D}_3 \otimes \dfrac{\partial \Delta \phi}{\partial \Delta \boldsymbol{\varepsilon}} ,
\end{equation}

In which
%\begin{align}
%\mathbb{D} &= 2\eta^2 K \mathbf{D}_1 \otimes \left[ -3\dfrac{\Delta \phi}{M^2} \mathbf{s}_n + \Delta \boldsymbol{\varepsilon}_d \right] + 2\eta \bar{\mu} \left( \mathbb{I} - \dfrac{1}{3} \mathbf{I} \otimes \mathbf{I} \right) \nonumber \\
%\mathbf{D}_3 &= 2\eta^2 K d_1 \left( -3\dfrac{\Delta \phi}{M^2} \mathbf{s}_n + \Delta \boldsymbol{\varepsilon}_d \right) - 6\dfrac{\eta^2 \bar{\mu}}{M^2} \left( \mathbf{s}_n + 2\bar{\mu} \Delta \boldsymbol{\varepsilon}_d \right) \nonumber
%\end{align}
%
\begin{align}
%\mathbb{D} &= \left( \mathbf{s}_n + 2\bar{\mu} \Delta \boldsymbol{\varepsilon}_d \right) \otimes \mathbf{D}_2 + 2 \eta \Delta \boldsymbol{\varepsilon}_d \otimes \mathbf{D}_1 + 2\eta \bar{\mu} \left( \mathbb{I} - \dfrac{1}{3} \mathbf{I} \otimes \mathbf{I} \right) \nonumber \\
\mathbb{D} &= \mathbf{s}_n \otimes \mathbf{D}_2 + \Delta \boldsymbol{\varepsilon}_d \otimes 2\left( \bar{\mu} \mathbf{D}_2 + \eta \mathbf{D}_1 \right) + 2\eta \bar{\mu} \mathbb{I}_d , \nonumber \\
%\mathbf{D}_3 &= d_2 \left( \mathbf{s}_n + 2\bar{\mu} \Delta \boldsymbol{\varepsilon}_d \right) + 2\eta d_1 \Delta \boldsymbol{\varepsilon}_d ,
\mathbf{D} &= \mathbf{D}_3 = d_2 \mathbf{s}_n + 2\left( \bar{\mu} d_2 + \eta d_1 \right) \Delta \boldsymbol{\varepsilon}_d .
\end{align}

\section{Linearization of the sub-stepping scheme, modified Cam-Clay model}
\label{app:mcc_tangent_substepping}

Taking linearization of \eqref{eq:p_mcc} and \eqref{eq:pc_mcc} for the sub-step $(k+1)$, with the remark that linearization of ${}^k p$ and ${}^k p_c$ do not vanish, one obtains:
\begin{equation}
\dfrac{\partial \, {}^{k+1}p}{\partial \Delta \boldsymbol{\varepsilon}} = \dfrac{{}^{k+1}p}{{}^{k}p} \dfrac{\partial \, {}^{k}p}{\partial \Delta \boldsymbol{\varepsilon}} + \dfrac{1+e}{\kappa} {}^{k+1} p \dfrac{\partial \, {}^{k+1}\Delta \varepsilon^e_v}{\partial \Delta \boldsymbol{\varepsilon}} ,
\label{eq:p_de_1_s}
\end{equation}
and
\begin{equation}
\dfrac{\partial \, {}^{k+1} p_c}{\partial \Delta \boldsymbol{\varepsilon}} = \dfrac{{}^{k+1} p_c}{{}^{k} p_c} \dfrac{\partial \, {}^{k} p_c}{\partial \Delta \boldsymbol{\varepsilon}} + \theta \, {}^{k+1} p_c \left[ \Delta \phi \left( 2\dfrac{\partial \, {}^{k+1} p}{\partial \Delta \boldsymbol{\varepsilon}} - \dfrac{\partial \, {}^{k+1} p_c}{\partial \Delta \boldsymbol{\varepsilon}} \right) + \left( 2 \, {}^{k+1} p - {}^{k+1} p_c \right) \dfrac{\partial \Delta \phi}{\partial \Delta \boldsymbol{\varepsilon}}  \right] .
\label{eq:pc_de_1_s}
\end{equation}

It is noted that ${}^{k+1} \Delta \varepsilon^e_v = {}^{k+1} \Delta \epsilon_v - \Delta \phi (2 \, {}^{k+1} p - {}^{k+1} p_c)$, hence:
\begin{equation}
\dfrac{\partial \, {}^{k+1} \Delta \varepsilon^e_v}{\partial \Delta \boldsymbol{\varepsilon}} = \alpha_{k+1} \mathbf{I} - \Delta \phi \left( 2 \dfrac{\partial \, {}^{k+1} p}{\partial \Delta \boldsymbol{\varepsilon}} - \dfrac{\partial \, {}^{k+1} p_c}{\partial \Delta \boldsymbol{\varepsilon}} \right) - \left( 2 \, {}^{k+1} p -  {}^{k+1} p_c \right) \dfrac{\partial \Delta \phi}{\partial \Delta \boldsymbol{\varepsilon}} .
\label{eq:de_e_v_de_1_s}
\end{equation}

Using \eqref{eq:de_e_v_de_1_s} and solve \eqref{eq:p_de_1_s} and \eqref{eq:pc_de_1_s} with respect to variable $\dfrac{\partial \, {}^{k+1}p}{\partial \Delta \boldsymbol{\varepsilon}}$ and $\dfrac{\partial \, {}^{k+1}p_c}{\partial \Delta \boldsymbol{\varepsilon}}$ leads to:
\begin{equation}
\dfrac{\partial \, {}^{k+1}p}{\partial \Delta \boldsymbol{\varepsilon}} = \mathbf{A}_1 + a_1 \dfrac{\partial \Delta \phi}{\partial \Delta \boldsymbol{\varepsilon}} \, ,
\qquad
\dfrac{\partial \, {}^{k+1}p_c}{\partial \Delta \boldsymbol{\varepsilon}} = \mathbf{B}_1 + b_1 \dfrac{\partial \Delta \phi}{\partial \Delta \boldsymbol{\varepsilon}} .
\label{eq:p_de_2_s_and_pc_de_2_s}
\end{equation}

In which
\begin{align}
\mathbf{A} &= \mathbf{A}_1 = \dfrac{1 + \theta \, {}^{k+1} p_c \Delta \phi}{a_2} \left( \dfrac{{}^{k+1} p}{{}^k p} \dfrac{\partial \, {}^{k} p}{\partial \Delta \boldsymbol{\varepsilon}} + \alpha_{k+1} {}^{k+1}K \mathbf{I} \right) + \dfrac{{}^{k+1} K \Delta \phi}{a_2} \dfrac{{}^{k+1} p_c}{{}^k p_c} \dfrac{\partial \, {}^{k} p_c}{\Delta \boldsymbol{\varepsilon}} , \nonumber
\\
a &= a_1 = - \dfrac{2 \, {}^{k+1} p - {}^{k+1} p_c}{a_2} {}^{k+1} K , \nonumber
\\
a_2 &= 1 + \left( \theta \, {}^{k+1} p_c + 2 \, {}^{k+1} K \right) \Delta \phi .
\\
%\mathbf{B} &= \mathbf{B}_1 = \dfrac{1}{1 + \theta \, {}^{k+1} p_c \Delta \phi} \left( \dfrac{{}^{k+1} p_c}{{}^k p_c} \dfrac{\partial \, {}^{k} p_c}{\partial \Delta \boldsymbol{\varepsilon}} + 2 \theta \, {}^{k+1} p_c \Delta \phi \, \mathbf{A}_1 \right) \nonumber
\mathbf{B} &= \mathbf{B}_1 = \dfrac{2 \theta \, {}^{k+1} p_c \Delta \phi}{a_2} \left( \dfrac{{}^{k+1} p}{{}^k p} \dfrac{\partial \, {}^{k} p}{\partial \Delta \boldsymbol{\varepsilon}} + \alpha_{k+1} {}^{k+1}K \mathbf{I} \right) + \dfrac{1 + 2\, {}^{k+1} K \Delta \phi}{a_2} \dfrac{{}^{k+1} p_c}{{}^k p_c} \dfrac{\partial \, {}^{k} p_c}{\Delta \boldsymbol{\varepsilon}} ,
\\
b &= b_1 = \dfrac{\theta \, {}^{k+1} p_c}{a_2} \left( 2 \, {}^{k+1} p - {}^{k+1} p_c \right) ,
\\
{}^{k+1} K &= \dfrac{1 + e}{\kappa} {}^{k+1} p .
\end{align}

From \Cref{eq:p_de_1_s}:
\begin{equation}
\dfrac{\partial \, {}^{k+1}\Delta \varepsilon^e_v}{\partial \Delta \boldsymbol{\varepsilon}} = \dfrac{1}{{}^{k+1} K} \left( \dfrac{\partial \, {}^{k+1} p}{\partial \Delta \boldsymbol{\varepsilon}} - \dfrac{{}^{k+1} p}{{}^{k} p} \dfrac{\partial \, {}^{k} p}{\partial \Delta \boldsymbol{\varepsilon}} \right) = \mathbf{C}_1 + c_1 \dfrac{\partial \Delta \phi}{\partial \Delta \boldsymbol{\varepsilon}} \, ,
\quad
\mathbf{C}_1 = \dfrac{1}{{}^{k+1}K} \left( \mathbf{A}_1 - \dfrac{{}^{k+1} p}{{}^{k} p} \dfrac{\partial \, {}^{k} p}{\partial \Delta \boldsymbol{\varepsilon}} \right) \, ,
\quad
c_1 = \dfrac{a_1}{{}^{k+1}K}
\label{eq:de_e_v_de_2_s}
\end{equation}

The deviatoric pressure reads:
\begin{equation}
{}^{k+1} \mathbf{s} = \eta \left( {}^k \mathbf{s} + 2 \bar{\mu} \, {}^{k+1} \Delta \boldsymbol{\varepsilon}_d \right) \, ,
\qquad
\bar{\mu} = \dfrac{{}^{k+1} p - {}^k p}{{}^{k+1} \, \Delta \epsilon_v^e} r \, ,
\qquad
\eta = \left( 1 + 6\bar{\mu}\dfrac{\Delta \phi}{M^2} \right)^{-1} .
\end{equation}

It is therefore:
\begin{equation}
\dfrac{\partial \bar{\mu}}{\partial \Delta \boldsymbol{\varepsilon}} = \dfrac{r}{{}^{k+1} \, \Delta \epsilon_v^e} \left( \dfrac{\partial \, {}^{k+1}p}{\partial \Delta \boldsymbol{\varepsilon}} - \dfrac{\partial \, {}^{k}p}{\partial \Delta \boldsymbol{\varepsilon}} \right) - \dfrac{\bar{\mu}}{{}^{k+1} \Delta \epsilon_v^e} \dfrac{\partial \, {}^{k+1}\Delta \varepsilon^e_v}{\partial \Delta \boldsymbol{\varepsilon}} .
\label{eq:mu_bar_de_1_s}
\end{equation}

Replace \eqref{eq:p_de_1_s} and \eqref{eq:de_e_v_de_2_s} into \eqref{eq:mu_bar_de_1_s}, one obtains:
\begin{equation}
\dfrac{\partial \bar{\mu}}{\partial \Delta \boldsymbol{\varepsilon}} = \mathbf{D}_1 + d_1 \dfrac{\partial \Delta \phi}{\partial \Delta \boldsymbol{\varepsilon}} \, ,
\qquad
\mathbf{D}_1 = \dfrac{r \mathbf{A}_1 - \bar{\mu}\mathbf{C}_1}{{}^{k+1} \Delta \varepsilon^e_v} - \dfrac{r}{{}^{k+1} \Delta \varepsilon^e_v} \dfrac{\partial \, {}^{k} p}{\partial \Delta \boldsymbol{\varepsilon}} \, ,
\qquad
d_1 = \dfrac{r a_1 - \bar{\mu}c_1}{{}^{k+1}\Delta \varepsilon^e_v} .
\end{equation}

The derivation of $\dfrac{\partial \eta}{\partial \Delta \boldsymbol{\varepsilon}}$ is similar to \eqref{eq:eta_de}, which reads:
\begin{equation}
\dfrac{\partial \eta}{\partial \Delta \boldsymbol{\varepsilon}} = -\dfrac{6 \eta^2}{M^2} \left( \Delta \phi \dfrac{\partial \bar{\mu}}{\partial \Delta \boldsymbol{\varepsilon}} + \bar{\mu} \dfrac{\partial \Delta \phi}{\partial \Delta \boldsymbol{\varepsilon}} \right) = \mathbf{D}_2 + d_2 \dfrac{\partial \Delta \phi}{\partial \Delta \boldsymbol{\varepsilon}} \, ,
\quad
\mathbf{D}_2 = -\dfrac{6 \eta^2 \Delta \phi}{M^2} \mathbf{D}_1 \, ,
\qquad
d_2 = -\dfrac{6\eta^2}{M^2} \left( \Delta \phi d_1 + \bar{\mu} \right) .
\end{equation}

The linearization of deviatoric pressure ${}^{k+1} \mathbf{s}$ follows by:
\begin{equation}
\dfrac{\partial \, {}^{k+1} \mathbf{s}}{\partial \Delta \boldsymbol{\varepsilon}} = \left( {}^k \mathbf{s} + 2 \bar{\mu} \, {}^{k+1} \Delta \boldsymbol{\varepsilon}_d \right) \otimes \dfrac{\partial \eta}{\partial \Delta \boldsymbol{\varepsilon}} + \eta \dfrac{\partial \, {}^k \mathbf{s}}{\partial \Delta \boldsymbol{\varepsilon}} + 2 \eta \, {}^{k+1} \Delta \boldsymbol{\varepsilon}_d \otimes \dfrac{\partial \bar{\mu}}{\partial \Delta \boldsymbol{\varepsilon}} + 2 \eta \bar{\mu} \dfrac{\partial \, {}^{k+1} \Delta \boldsymbol{\varepsilon}_d}{\partial \Delta \boldsymbol{\varepsilon}} = \mathbb{D} + \mathbf{D}_3 \otimes \dfrac{\Delta \phi}{\Delta \boldsymbol{\varepsilon}} ,
\label{eq:s_de_1_s}
\end{equation}

In which
\begin{align}
\mathbb{D} &= \left( {}^k\mathbf{s} + 2 \bar{\mu} \, {}^{k+1} \Delta \boldsymbol{\varepsilon}_d \right) \otimes \mathbf{D}_2 + \eta \dfrac{\partial \, {}^k \mathbf{s}}{\partial \Delta \boldsymbol{\varepsilon}} + 2 \eta \, {}^{k+1} \Delta \boldsymbol{\varepsilon}_d \otimes \mathbf{D}_1 + 2 \alpha_{k+1} \eta \bar{\mu} \mathbb{I}_d , \nonumber \\
\mathbf{D} &= \mathbf{D}_3 = d_2 \left( {}^k\mathbf{s} + 2 \bar{\mu} \, {}^{k+1} \Delta \boldsymbol{\varepsilon}_d \right) + 2 \eta d_1 
\, {}^{k+1} \Delta \boldsymbol{\varepsilon}_d .
\end{align}

\section{Solution of the local nonlinear system, CASM model}
\label{app:casm_local_solution}

In order to derive the solution of the local nonlinear system, necessary derivatives of the yield function and plastic potential are given to support for further derivation:
\begin{equation}
f_{p} = \dfrac{\partial f}{\partial p} = - \dfrac{N q^N}{M^N p^{N+1}} + \dfrac{1}{p \ln R} \, ,
\qquad
f_{q} = \dfrac{\partial f}{\partial q} = \dfrac{N q^{N-1}}{M^N p^N} \, ,
\qquad
f_{p_c} = \dfrac{\partial f}{\partial p_c} = - \dfrac{1}{p_c \ln R} \, ,
\end{equation}
\begin{equation}
g_p = \dfrac{\partial g}{\partial p} = 3 \dfrac{3+2M}{3p + 2q} - 3 \dfrac{3-M}{3p - q} \, ,
\qquad
g_q = \dfrac{\partial g}{\partial q} = 2 \dfrac{3+2M}{3p+2q} + \dfrac{3-M}{3p - q} \, ,
\end{equation}
\begin{equation}
g_{pp} = \dfrac{\partial^2 g}{\partial p^2} = -9 \dfrac{3+2M}{(3p+2q)^2} + 9 \dfrac{3-M}{(3p-q)^2} \, ,
\quad
g_{qq} = \dfrac{\partial^2 g}{\partial q^2} = -4 \dfrac{3+2M}{(3p+2q)^2} + \dfrac{3-M}{(3p-q)^2} \, ,
\end{equation}
\begin{equation}
g_{pq} = \dfrac{\partial^2 q}{\partial p \partial q} = -6 \dfrac{3+2M}{(3p+2q)^2} - 3 \dfrac{3-M}{(3p-q)^2} .
\end{equation}

Additional terms are defined as:
\begin{equation}
K = \dfrac{1+e}{\kappa} p \, ,
\qquad
\bar{\mu} = K r \, ,
\qquad
\eta = \left( 1 + 3\bar{\mu} \dfrac{\Delta \phi}{q} g_q \right)^{-1} \, ,
\qquad
\hat{\mathbf{n}} = \dfrac{\mathbf{s}_n + 2\bar{\mu} \Delta \bs{\varepsilon}_d}{\| \mathbf{s}_n + 2\bar{\mu} \Delta \bs{\varepsilon}_d \|} .
\end{equation}

The flow rule reads:
\begin{equation}
\Delta \varepsilon^p_v = \Delta \phi \, g_p \, ,
\qquad
\Delta \varepsilon^e_v = \Delta \epsilon_v - \Delta \varepsilon^p_v \, ,
\qquad
\Delta \bs{\varepsilon}^p_d = \sqrt{\dfrac{3}{2}} \Delta \phi \, g_q \hat{\mathbf{n}} \, ,
\qquad
\Delta \bs{\varepsilon}^e_d = \Delta \bs{\varepsilon}_d - \Delta \bs{\varepsilon}^p_d .
\end{equation}

The derivatives of $\bar{\mu}$ and $\eta$ read:
\begin{equation}
\bar{\mu}_p = \dfrac{\partial \bar{\mu}}{\partial p} = \dfrac{1}{\Delta \varepsilon^e_v} \left( r + \bar{\mu} \Delta \phi \, g_{pp} \right) \, ,
\quad
\bar{\mu}_q = \dfrac{\partial \bar{\mu}}{\partial q} = \dfrac{\bar{\mu}}{\Delta \varepsilon^e_v} \Delta \phi \, g_{pq} \,
\quad
\bar{\mu}_{p_c} = \dfrac{\partial \bar{\mu}}{\partial p_c} = 0 \, ,
\quad
\bar{\mu}_{\Delta \phi} = \dfrac{\partial \bar{\mu}}{\partial \Delta \phi} = \dfrac{\bar{\mu}}{\Delta \varepsilon^e_v} g_{p} \, ,
\end{equation}
\begin{equation}
\begin{split}
\eta_p = \dfrac{\partial \eta}{\partial p} = - 3 \dfrac{\Delta \phi}{q} \left( \bar{\mu}_p g_q + \bar{\mu} g_{pq} \right) \eta^2 \, &,
\quad
\eta_q = \dfrac{\partial \eta}{\partial p} = - 3 \dfrac{\Delta \phi}{q} \left( \left( \bar{\mu}_q - \dfrac{\bar{\mu}}{q} \right) g_q + \bar{\mu} g_{qq} \right) \eta^2 \, , \\
\bar{\eta}_{p_c} = \dfrac{\partial \eta}{\partial p_c} = 0 \, &,
\quad
\bar{\eta}_{\Delta \phi} = \dfrac{\partial \eta}{\partial \Delta \phi} = - \dfrac{3}{q} \left( \bar{\mu} + \bar{\mu}_{\Delta \phi} \Delta \phi \right) g_q \eta^2 .
\end{split}
\end{equation}

The necessary derivatives to solver local system \eqref{eq:p_mcc} \eqref{eq:q_mcc} \eqref{eq:pc_mcc} \eqref{eq:f_mcc} are given as follow:
\begin{equation}
h_{1,1} = \dfrac{\partial h_1}{\partial p} = 1 + K \Delta \phi \, g_{pp} \, ,
\quad
h_{1,2} = \dfrac{\partial h_1}{\partial q} = K \Delta \phi \, g_{pq} \, ,
\quad
h_{1,3} = \dfrac{\partial h_1}{\partial p_c} = 0 \, ,
\quad
h_{1,4} = \dfrac{\partial h_1}{\partial \Delta \phi} =  K g_{p} .
\nonumber
\end{equation}
\begin{equation}
\begin{split}
h_{2,1} = \dfrac{\partial h_2}{\partial p} = - \sqrt{\dfrac{3}{2}} \eta_p \| \mathbf{s}_n + 2\bar{\mu} \Delta \bs{\varepsilon}_d \| - 2 \sqrt{\dfrac{3}{2}} \eta \bar{\mu}_p \hat{\mathbf{n}} : \Delta \bs{\varepsilon}_d \, &,
\quad
h_{2,2} = \dfrac{\partial h_2}{\partial q} = 1 - \sqrt{\dfrac{3}{2}} \eta_q \| \mathbf{s}_n + 2\bar{\mu} \Delta \bs{\varepsilon}_d \| - 2 \sqrt{\dfrac{3}{2}} \eta \bar{\mu}_q \hat{\mathbf{n}} : \Delta \bs{\varepsilon}_d \, , \\
h_{2,3} = \dfrac{\partial h_2}{\partial p_c} = - 2 \sqrt{\dfrac{3}{2}} \eta \bar{\mu}_{p_c} \hat{\mathbf{n}} : \Delta \bs{\varepsilon}_d \, &,
\quad
h_{2,4} = \dfrac{\partial h_2}{\partial \Delta \phi} = - \sqrt{\dfrac{3}{2}} \eta_{\Delta \phi} \| \mathbf{s}_n + 2\bar{\mu} \Delta \bs{\varepsilon}_d \| - 2 \sqrt{\dfrac{3}{2}} \eta \bar{\mu}_{\Delta \phi} \hat{\mathbf{n}} : \Delta \bs{\varepsilon}_d \, ,
\end{split}
\nonumber
\end{equation}
\begin{equation}
h_{3,1} = \dfrac{\partial h_3}{\partial p} = - \theta p_c \Delta \phi \, g_{pp} \, ,
\quad
h_{3,2} = \dfrac{\partial h_3}{\partial q} = - \theta p_c \Delta \phi \, g_{pq} \, ,
\quad
h_{3,3} = \dfrac{\partial h_3}{\partial p_c} = 1 \, ,
\quad
h_{3,4} = \dfrac{\partial h_3}{\partial \Delta \phi} = - \theta p_c \, g_{p} \, ,
\nonumber
\end{equation}
\begin{equation}
h_{4,1} = \dfrac{\partial h_4}{\partial p} = f_p \, ,
\quad
h_{4,2} = \dfrac{\partial h_4}{\partial q} = f_q \, ,
\quad
h_{4,3} = \dfrac{\partial h_4}{\partial p_c} = f_{p_c} \, ,
\quad
h_{4,4} = \dfrac{\partial h_4}{\partial \Delta \phi} = 0 .
\nonumber
\end{equation}

\section{Linearization of the one-step scheme, CASM model}
\label{app:casm_tangent}

Taking the linearization of \eqref{eq:p_mcc}, one obtains:
\begin{equation}
\dfrac{\partial p}{\partial \Delta \boldsymbol{\varepsilon}} = K \dfrac{\partial \Delta \varepsilon^e_v}{\partial \Delta \boldsymbol{\varepsilon}} \, ,
\qquad
K = \dfrac{1+e}{\kappa} p .
\label{eq:p_de_1_casm}
\end{equation}

It is noted that $\Delta \varepsilon^e_v = \Delta \epsilon_v - \Delta \phi g_p$, hence:
\begin{equation}
\dfrac{\partial \Delta \varepsilon^e_v}{\partial \Delta \boldsymbol{\varepsilon}} = \mathbf{I} - \Delta \phi g_{pp} \dfrac{\partial  p}{\partial \Delta \bs{\varepsilon}} - \Delta \phi g_{pq} \dfrac{\partial  q}{\partial \Delta \bs{\varepsilon}} - g_p \dfrac{\partial \Delta \phi}{\partial \Delta \boldsymbol{\varepsilon}} .
\label{eq:de_e_v_de_1_casm}
\end{equation}

From \eqref{eq:p_de_1_casm} and \eqref{eq:de_e_v_de_1_casm}, one can rewrite $\dfrac{\partial p}{\partial \Delta \boldsymbol{\varepsilon}}$ as dependent term of $\dfrac{\partial q}{\partial \Delta \boldsymbol{\varepsilon}}$ and $\dfrac{\partial \Delta \phi}{\partial \Delta \boldsymbol{\varepsilon}}$:
\begin{equation}
\begin{gathered}
\dfrac{\partial p}{\partial \Delta \boldsymbol{\varepsilon}} = \mathbf{A}_1 + a_1 \dfrac{\partial \Delta \phi}{\partial \Delta \boldsymbol{\varepsilon}} + a_2 \dfrac{\partial q}{\partial \Delta \boldsymbol{\varepsilon}} \, ,
\quad
\mathbf{A}_1 = a_3 K \mathbf{I} \, , \\
a_1 = -a_3 K g_p \, ,
\quad
a_2 = -a_3 K \Delta \phi g_{pq} \, ,
\quad
a_3 = \left( 1 + K \Delta \phi g_{pp} \right)^{-1} .
\end{gathered}
\label{eq:p_de_casm}
\end{equation}

Taking the linearization of \eqref{eq:pc_mcc}:
\begin{equation}
\begin{gathered}
\dfrac{\partial p_c}{\partial \Delta \boldsymbol{\varepsilon}} = \theta p_c \left( \mathbf{I} - \dfrac{\partial \Delta \varepsilon^e_v}{\partial \Delta \boldsymbol{\varepsilon}} \right) = \mathbf{B}_1 + b_1 \dfrac{\partial \Delta \phi}{\partial \Delta \boldsymbol{\varepsilon}} + b_2 \dfrac{\partial q}{\partial \Delta \boldsymbol{\varepsilon}} \, , \\
\mathbf{B}_1 = \dfrac{\theta p_c}{K} \left( K \mathbf{I} - \mathbf{A}_1 \right) \, ,
\quad
b_1 = -\dfrac{\theta p_c}{K} a_1 \, ,
\quad
b_2 = -\dfrac{\theta p_c}{K} a_2 .
\end{gathered}
\label{eq:pc_de_casm}
\end{equation}

The deviatoric pressure reads:
\begin{equation}
\mathbf{s} = \eta \left( \mathbf{s}_n + 2 \bar{\mu} \Delta \boldsymbol{\varepsilon}_d \right) \, ,
\qquad
\bar{\mu} = \dfrac{p - p_n}{\Delta \epsilon_v^e} r \, ,
\qquad
\eta = \left( 1 + 3 \bar{\mu}\dfrac{\Delta \phi}{q} g_{q} \right)^{-1} .
\end{equation}

The linearization of $\bar{\mu}$ comes as follows:
\begin{equation}
\begin{gathered}
\dfrac{\partial \bar{\mu}}{\partial \Delta \bs{\varepsilon}} = \mathbf{D}_1 + d_1 \dfrac{\partial \Delta \phi}{\partial \Delta \boldsymbol{\varepsilon}} + d_2 \dfrac{\partial q}{\partial \Delta \boldsymbol{\varepsilon}} \, ,
\quad
\mathbf{D}_1 = \dfrac{K r - \bar{\mu}}{K \Delta \epsilon_v^e} \mathbf{A}_1 \, ,
\\
d_1 = \dfrac{K r - \bar{\mu}}{K \Delta \epsilon_v^e} a_1 \, ,
\qquad
d_2 = \dfrac{K r - \bar{\mu}}{K \Delta \epsilon_v^e} a_2 .
\end{gathered}
\end{equation}

The linearization of $\eta$ comes as follows:
\begin{equation}
\begin{gathered}
\dfrac{\partial \eta}{\partial \Delta \bs{\varepsilon}} = \mathbf{D}_2 + d_3 \dfrac{\partial \Delta \phi}{\partial \Delta \boldsymbol{\varepsilon}} + d_4 \dfrac{\partial q}{\partial \Delta \boldsymbol{\varepsilon}} \, ,
\quad
\mathbf{D}_2 = -3 \eta^2 \dfrac{\Delta \phi}{q} \left( g_q \mathbf{D}_1 + \bar{\mu} g_{pq} \mathbf{A}_1 \right) \, ,
\\
d_3 = -3 \eta^2 \left( \left[ \dfrac{\Delta \phi}{q} d_1 + \dfrac{\bar{\mu}}{q} \right] g_q + \dfrac{\bar{\mu} \Delta \phi}{q} g_{pq} a_1 \right) \, ,
\quad
d_4 = -3 \eta^2 \dfrac{\Delta \phi}{q} \left( \left[ d_2 - \dfrac{\bar{\mu}}{q} \right] g_q + \bar{\mu} [ g_{pq} a_2 + g_{qq} ] \right) .
\end{gathered}
\end{equation}

The linearization of $\mathbf{s}$ comes as follows:
\begin{equation}
\begin{gathered}
\dfrac{\partial \mathbf{s}}{\partial \Delta \bs{\varepsilon}} = 2 \eta \left( \bar{\mu} \mathbb{I}_d + \Delta \bs{\varepsilon}_d \otimes \dfrac{\partial \bar{\mu}}{\partial \Delta \bs{\varepsilon}} \right) + \left( \mathbf{s}_n + 2 \bar{\mu} \Delta \bs{\varepsilon}_d \right) \otimes \dfrac{\partial \eta}{\partial \Delta \bs{\varepsilon}} = \mathbb{D}_1 + \mathbf{D}_3 \otimes \dfrac{\partial \Delta \phi}{\partial \Delta \bs{\varepsilon}} + \mathbf{D}_4 \otimes \dfrac{\partial q}{\partial \Delta \bs{\varepsilon}} \, ,
\\
\mathbb{D}_1 = 2 \eta \bar{\mu} \mathbb{I}_d + 2 \eta \Delta \bs{\varepsilon}_d \otimes \mathbf{D}_1 + \left( \mathbf{s}_n + 2 \bar{\mu} \Delta \bs{\varepsilon}_d \right) \otimes \mathbf{D}_2 \, ,
\\
\mathbf{D}_3 = d_3 \mathbf{s}_n + 2 \left( \bar{\mu} d_3 + \eta d_1 \right) \Delta \bs{\varepsilon}_d \, ,
\quad
\mathbf{D}_4 = d_4 \mathbf{s}_n + 2 \left( \bar{\mu} d_4 + \eta d_2 \right) \Delta \bs{\varepsilon}_d .
\end{gathered}
\label{eq:s_de_casm}
\end{equation}

From the relation $2 q \dfrac{\partial q}{\partial \Delta \bs{\varepsilon}} = 3 \mathbf{s} : \dfrac{\partial \mathbf{s}}{\partial \Delta \bs{\varepsilon}}$, the linearization of $q$ reduces to:
\begin{equation}
\dfrac{\partial q}{\partial \Delta \bs{\varepsilon}} = \mathbf{D}_5 + d_5 \dfrac{\partial \Delta \phi}{\partial \Delta \bs{\varepsilon}} \, ,
\quad
\mathbf{D}_5 = \dfrac{3}{2} \dfrac{d_6}{q} \mathbf{s} : \mathbb{D}_1 \, ,
\quad
d_5 = \dfrac{3}{2} \dfrac{d_6}{q} \mathbf{s} : \mathbf{D}_3 \, ,
\quad
d_6 = \left( 1 - \dfrac{3}{2} \dfrac{1}{q} \mathbf{s} : \mathbf{D}_4 \right)^{-1} .
\label{eq:q_de_casm}
\end{equation}

Replacing \eqref{eq:q_de_casm} back into \eqref{eq:p_de_casm}, \eqref{eq:pc_de_casm} and \eqref{eq:s_de_casm}, we obtain the dependency of linearization of $p$, $p_c$ and $\mathbf{s}$ on only $\dfrac{\partial \Delta \phi}{\partial \Delta \bs{\varepsilon}}$:
\begin{equation}
\begin{gathered}
\dfrac{\partial p}{\partial \Delta \bs{\varepsilon}} = \mathbf{A} + a \dfrac{\partial \Delta \phi}{\partial \Delta \bs{\varepsilon}} \, ,
\qquad
\mathbf{A} = \mathbf{A}_1 + a_2 \mathbf{D}_5 \, ,
\qquad
a = a_1 + a_2 d_5 \, ,
\\
\dfrac{\partial p_c}{\partial \Delta \bs{\varepsilon}} = \mathbf{B} + b \dfrac{\partial \Delta \phi}{\partial \Delta \bs{\varepsilon}} \, ,
\qquad
\mathbf{B} = \mathbf{B}_1 + b_2 \mathbf{D}_5 \, ,
\qquad
b = b_1 + b_2 d_5 \, ,
\\
\dfrac{\partial \mathbf{s}}{\partial \Delta \bs{\varepsilon}} = \mathbb{D} + \mathbf{D} \dfrac{\partial \Delta \phi}{\partial \Delta \bs{\varepsilon}} \, ,
\qquad
\mathbb{D} = \mathbb{D}_1 + \mathbf{D}_4 \otimes \mathbf{D}_5 \, ,
\qquad
\mathbf{D} = \mathbf{D}_3 + d_5 \mathbf{D}_4 .
\end{gathered}
\end{equation}

\section{Linearization of the sub-stepping scheme, CASM model}
\label{app:casm_tangent_substepping}

Taking linearization of \eqref{eq:p_mcc} for the sub-step $(k+1)$, with the remark that linearization of ${}^k p$ does not vanish, one obtains:
\begin{equation}
\dfrac{\partial \, {}^{k+1}p}{\partial \Delta \boldsymbol{\varepsilon}} = \dfrac{{}^{k+1}p}{{}^{k}p} \dfrac{\partial \, {}^{k}p}{\partial \Delta \boldsymbol{\varepsilon}} + {}^{k+1} K \dfrac{\partial \, {}^{k+1}\Delta \varepsilon^e_v}{\partial \Delta \boldsymbol{\varepsilon}} \, ,
\qquad
{}^{k+1} K = \dfrac{1+e}{\kappa} {}^{k+1} p .
\label{eq:p_de_1_casm_s}
\end{equation}

It is noted that ${}^{k+1} \Delta \varepsilon^e_v = {}^{k+1} \Delta \epsilon_v - \Delta \phi g_p$, hence:
\begin{equation}
\dfrac{\partial \, {}^{k+1} \Delta \varepsilon^e_v}{\partial \Delta \boldsymbol{\varepsilon}} = \alpha_{k+1} \mathbf{I} - \Delta \phi \left( g_{pp} \dfrac{\partial \, {}^{k+1} p}{\partial \Delta \bs{\varepsilon}} + g_{pq} \dfrac{\partial \, {}^{k+1} q}{\partial \Delta \bs{\varepsilon}} \right) - g_p \dfrac{\partial \Delta \phi}{\partial \Delta \boldsymbol{\varepsilon}} .
\label{eq:de_e_v_de_1_casm_s}
\end{equation}

In \Cref{eq:de_e_v_de_1_casm_s} and further equations below, $g_p$, $g_q$, $g_{pp}$, $g_{pq}$ and $g_{qq}$ are evaluated at $({}^{k+1} p, {}^{k+1} q)$.

From \eqref{eq:p_de_1_casm_s} and \eqref{eq:de_e_v_de_1_casm_s}, one can rewrite $\dfrac{\partial \, {}^{k+1} p}{\partial \Delta \boldsymbol{\varepsilon}}$ as dependent term of $\dfrac{\partial \, {}^{k+1} q}{\partial \Delta \boldsymbol{\varepsilon}}$ and $\dfrac{\partial \Delta \phi}{\partial \Delta \boldsymbol{\varepsilon}}$:
\begin{equation}
\begin{gathered}
\dfrac{\partial \, {}^{k+1} p}{\partial \Delta \boldsymbol{\varepsilon}} = \mathbf{A}_1 + a_1 \dfrac{\partial \Delta \phi}{\partial \Delta \boldsymbol{\varepsilon}} + a_2 \dfrac{\partial \, {}^{k+1} q}{\partial \Delta \boldsymbol{\varepsilon}} \, ,
\quad
\mathbf{A}_1 = a_3 \left( \dfrac{{}^{k+1}p}{{}^{k}p} \dfrac{\partial \, {}^{k}p}{\partial \Delta \boldsymbol{\varepsilon}} + \alpha_{k+1} \, {}^{k+1} K \mathbf{I} \right) \, , \\
a_1 = -a_3 {}^{k+1} K g_p \, ,
\quad
a_2 = -a_3 {}^{k+1} K \Delta \phi g_{pq} \, ,
\quad
a_3 = \left( 1 + {}^{k+1} K \Delta \phi g_{pp} \right)^{-1} .
\end{gathered}
\label{eq:p_de_casm_s}
\end{equation}

Taking the linearization of \eqref{eq:pc_mcc}:
\begin{equation}
\begin{gathered}
\dfrac{\partial \, {}^{k+1} p_c}{\partial \Delta \boldsymbol{\varepsilon}} = \dfrac{{}^{k+1} p_c}{{}^{k} p_c} \dfrac{\partial \, {}^{k} p_c}{\partial \Delta \boldsymbol{\varepsilon}} + \theta \, {}^{k+1} p_c \left( \alpha_{k+1} \mathbf{I} - \dfrac{\partial \, {}^{k+1} \Delta \varepsilon^e_v}{\partial \Delta \boldsymbol{\varepsilon}} \right) = \mathbf{B}_1 + b_1 \dfrac{\partial \Delta \phi}{\partial \Delta \boldsymbol{\varepsilon}} + b_2 \dfrac{\partial \, {}^{k+1} q}{\partial \Delta \boldsymbol{\varepsilon}} \, , \\
\mathbf{B}_1 = \dfrac{{}^{k+1} p_c}{{}^{k} p_c} \dfrac{\partial \, {}^{k} p_c}{\partial \Delta \boldsymbol{\varepsilon}} + \dfrac{\theta \, {}^{k+1} p_c}{{}^{k+1} K} \left( \alpha_{k+1} {}^{k+1} K \mathbf{I} + \dfrac{{}^{k+1} p}{{}^{k} p} \dfrac{\partial \, {}^{k} p}{\partial \Delta \bs{\varepsilon}} - \mathbf{A}_1 \right) \, ,
\quad
b_1 = -\dfrac{\theta \, {}^{k+1} p_c}{{}^{k+1} K} a_1 \, ,
\quad
b_2 = -\dfrac{\theta \, {}^{k+1} p_c}{{}^{k+1} K} a_2 .
\end{gathered}
\label{eq:pc_de_casm_s}
\end{equation}

The deviatoric pressure reads:
\begin{equation}
{}^{k+1} \mathbf{s} = \eta \left( {}^k \mathbf{s} + 2 \bar{\mu} \, {}^{k+1}\Delta \boldsymbol{\varepsilon}_d \right) \, ,
\qquad
\bar{\mu} = \dfrac{{}^{k+1} p - {}^k p}{{}^{k+1} \, \Delta \epsilon_v^e} r \, ,
\qquad
\eta = \left( 1 + 3 \bar{\mu} \dfrac{\Delta \phi}{{}^{k+1} q} g_q \right)^{-1} .
\end{equation}

The linearization of $\bar{\mu}$ comes as follows:
\begin{equation}
\begin{gathered}
\dfrac{\partial \bar{\mu}}{\partial \Delta \bs{\varepsilon}} = \mathbf{D}_1 + d_1 \dfrac{\partial \Delta \phi}{\partial \Delta \boldsymbol{\varepsilon}} + d_2 \dfrac{\partial \, {}^{k+1} q}{\partial \Delta \boldsymbol{\varepsilon}} \, ,
\quad
\mathbf{D}_1 = \dfrac{{}^{k+1} K r - \bar{\mu}}{{}^{k+1} K \, {}^{k+1} \Delta \epsilon_v^e} \mathbf{A}_1 - \dfrac{1}{^{k+1} \Delta \epsilon_v^e} \left( r - \dfrac{\bar{\mu}}{^{k+1} K} \dfrac{^{k+1} p}{^{k} p} \right) \dfrac{\partial \, ^{k} p}{\partial \Delta \bs{\varepsilon}} \, ,
\\
d_1 = \dfrac{{}^{k+1} K r - \bar{\mu}}{{}^{k+1} K \, {}^{k+1} \Delta \epsilon_v^e} a_1 \, ,
\qquad
d_2 = \dfrac{{}^{k+1} K r - \bar{\mu}}{{}^{k+1} K \, {}^{k+1} \Delta \epsilon_v^e} a_2 .
\end{gathered}
\end{equation}

The linearization of $\eta$ comes as follows:
\begin{equation}
\begin{gathered}
\dfrac{\partial \eta}{\partial \Delta \bs{\varepsilon}} = \mathbf{D}_2 + d_3 \dfrac{\partial \Delta \phi}{\partial \Delta \boldsymbol{\varepsilon}} + d_4 \dfrac{\partial \, {}^{k+1} q}{\partial \Delta \boldsymbol{\varepsilon}} \, ,
\quad
\mathbf{D}_2 = -3 \eta^2 \dfrac{\Delta \phi}{{}^{k+1} q} \left( g_q \mathbf{D}_1 + \bar{\mu} g_{pq} \mathbf{A}_1 \right) \, ,
\\
d_3 = -3 \eta^2 \left( \left[ \dfrac{\Delta \phi}{{}^{k+1} q} d_1 + \dfrac{\bar{\mu}}{{}^{k+1} q} \right] g_q + \dfrac{\bar{\mu} \Delta \phi}{{}^{k+1} q} g_{pq} a_1 \right) \, ,
\quad
d_4 = -3 \eta^2 \dfrac{\Delta \phi}{{}^{k+1} q} \left( \left[ d_2 - \dfrac{\bar{\mu}}{{}^{k+1} q} \right] g_q + \bar{\mu} [ g_{pq} a_2 + g_{qq} ] \right) .
\end{gathered}
\end{equation}

The linearization of ${}^{k+1} \mathbf{s}$ comes as follows:
\begin{equation}
\begin{gathered}
\dfrac{\partial \, {}^{k+1} \mathbf{s}}{\partial \Delta \bs{\varepsilon}} = \eta \left( \dfrac{\partial \, {}^{k} \mathbf{s}}{\partial \Delta \bs{\varepsilon}} + 2 \alpha_{k+1} \bar{\mu} \mathbb{I}_d + 2 \, {}^{k+1} \Delta \bs{\varepsilon}_d \otimes \dfrac{\partial \bar{\mu}}{\partial \Delta \bs{\varepsilon}} \right) + \left( {}^{k} \mathbf{s} + 2 \bar{\mu} \, {}^{k+1} \Delta \bs{\varepsilon}_d \right) \otimes \dfrac{\partial \eta}{\partial \Delta \bs{\varepsilon}} = \mathbb{D}_1 + \mathbf{D}_3 \otimes \dfrac{\partial \Delta \phi}{\partial \Delta \bs{\varepsilon}} + \mathbf{D}_4 \otimes \dfrac{\partial \, {}^{k+1} q}{\partial \Delta \bs{\varepsilon}} \, ,
\\
\mathbb{D}_1 = \eta \left( \dfrac{\partial \, {}^{k} \mathbf{s}}{\partial \Delta \bs{\varepsilon}} + 2 \, \alpha_{k+1} \bar{\mu} \mathbb{I}_d + 2 \, {}^{k+1} \Delta \bs{\varepsilon}_d \otimes \mathbf{D}_1 \right) + \left( {}^{k} \mathbf{s} + 2 \bar{\mu} \, {}^{k+1} \Delta \bs{\varepsilon}_d \right) \otimes \mathbf{D}_2 \, ,
\\
\mathbf{D}_3 = d_3 \, {}^k \mathbf{s} + 2 \left( \bar{\mu} d_3 + \eta d_1 \right) \, {}^{k+1} \Delta \bs{\varepsilon}_d \, ,
\quad
\mathbf{D}_4 = d_4 \, {}^k \mathbf{s} + 2 \left( \bar{\mu} d_4 + \eta d_2 \right) \, {}^{k+1} \Delta \bs{\varepsilon}_d .
\end{gathered}
\label{eq:s_de_casm_s}
\end{equation}

From the relation $2 \, {}^{k+1} q \dfrac{\partial \, {}^{k+1} q}{\partial \Delta \bs{\varepsilon}} = 3 \, {}^{k+1} \mathbf{s} : \dfrac{\partial \, {}^{k+1} \mathbf{s}}{\partial \Delta \bs{\varepsilon}}$, the linearization of ${}^{k+1} q$ reduces to:
\begin{equation}
\dfrac{\partial \, {}^{k+1} q}{\partial \Delta \bs{\varepsilon}} = \mathbf{D}_5 + d_5 \dfrac{\partial \Delta \phi}{\partial \Delta \bs{\varepsilon}} \, ,
\quad
\mathbf{D}_5 = \dfrac{3}{2} \dfrac{d_6}{{}^{k+1} q} {}^{k+1} \mathbf{s} : \mathbb{D}_1 \, ,
\quad
d_5 = \dfrac{3}{2} \dfrac{d_6}{{}^{k+1} q} {}^{k+1} \mathbf{s} : \mathbf{D}_3 \, ,
\quad
d_6 = \left( 1 - \dfrac{3}{2} \dfrac{1}{{}^{k+1} q} {}^{k+1} \mathbf{s} : \mathbf{D}_4 \right)^{-1} .
\label{eq:q_de_casm_s}
\end{equation}

Replacing \eqref{eq:q_de_casm_s} back into \eqref{eq:p_de_casm_s}, \eqref{eq:pc_de_casm_s} and \eqref{eq:s_de_casm_s}, we obtain the dependency of linearization of $p$, $p_c$ and $\mathbf{s}$ on only $\dfrac{\partial \Delta \phi}{\partial \Delta \bs{\varepsilon}}$:
\begin{equation}
\begin{gathered}
\dfrac{\partial \, {}^{k+1} p}{\partial \Delta \bs{\varepsilon}} = \mathbf{A} + a \dfrac{\partial \Delta \phi}{\partial \Delta \bs{\varepsilon}} \, ,
\qquad
\mathbf{A} = \mathbf{A}_1 + a_2 \mathbf{D}_5 \, ,
\qquad
a = a_1 + a_2 d_5 \, ,
\\
\dfrac{\partial \, {}^{k+1} p_c}{\partial \Delta \bs{\varepsilon}} = \mathbf{B} + b \dfrac{\partial \Delta \phi}{\partial \Delta \bs{\varepsilon}} \, ,
\qquad
\mathbf{B} = \mathbf{B}_1 + b_2 \mathbf{D}_5 \, ,
\qquad
b = b_1 + b_2 d_5 \, ,
\\
\dfrac{\partial \, {}^{k+1} \mathbf{s}}{\partial \Delta \bs{\varepsilon}} = \mathbb{D} + \mathbf{D} \dfrac{\partial \Delta \phi}{\partial \Delta \bs{\varepsilon}} \, ,
\qquad
\mathbb{D} = \mathbb{D}_1 + \mathbf{D}_4 \otimes \mathbf{D}_5 \, ,
\qquad
\mathbf{D} = \mathbf{D}_3 + d_5 \mathbf{D}_4 .
\end{gathered}
\end{equation}

\section{Analytical solution of CASM model subjected to drained proportional stress path}
\label{app:casm_load_sol}

In the proportional loading scheme, the loading is governed by:
\begin{equation}
\dot{q} = k \dot{p} \, , \qquad k = const .
\label{eq:proportional_loading_def}
\end{equation}

For further derivation, we denote $\beta = \dfrac{q}{p}$, subsequently:
\begin{equation}
q = \beta p \, ,
\qquad
\dot{q} = \dot{\beta} p + \beta \dot{p} .
\label{eq:loading_eta}
\end{equation}

Because of \Cref{eq:proportional_loading_def}, one can deduce
\begin{equation}
k \dot{p} = \dot{\beta} p + \beta \dot{p} \qquad \Rightarrow \qquad \dfrac{\dot{p}}{p} = \dfrac{\dot{\beta}}{k - \beta} .
\label{eq:casm_pld_p_dot_over_p}
\end{equation}

\subsection{Solution for volumetric strain}

\subsubsection{Elastic domain}

The \Cref{eq:elastic_mcc,eq:stiff_mcc} leads to:
\begin{equation}
\dot{\varepsilon}^e_v = \dfrac{\kappa}{1 + e} \dfrac{\dot{p}}{p} .
\label{eq:casm_pld_e_v_dot}
\end{equation}

In the elastic domain: $\dot{\varepsilon}_v = \dot{\varepsilon}^e_v$. Hence, $\Delta \epsilon_v = \Delta \varepsilon^e_v$ can be approximated by:
\begin{equation}
\Delta \varepsilon^e_v = \dfrac{\kappa}{1 + e} \Delta \ln p = \dfrac{\kappa}{1 + e} \ln \dfrac{p_{n+1}}{p_n} .
\label{eq:casm_pld_d_e_v}
\end{equation}

\subsubsection{Plastic domain}

In the plastic domain, $\Delta \varepsilon^e_v$ can be calculated using \Cref{eq:casm_pld_d_e_v}. To compute $\Delta {\varepsilon}^p_v$, we start by taking the time derivative of yield function \eqref{eq:yield_casm}:
\begin{equation}
\dfrac{\dot{p}_c}{p_c} = \dfrac{\dot{p}}{p} + N \ln R \dfrac{q^{N-1} \dot{q}}{(M p)^N} - N \ln R \, \dfrac{\dot{p}}{p} \left( \dfrac{q}{M p} \right)^N .
\label{eq:casm_pld_p_c_dot}
\end{equation}

Taking into account \Cref{eq:proportional_loading_def,eq:casm_pld_p_dot_over_p}, \Cref{eq:casm_pld_p_c_dot} can be rewritten as:
\begin{equation}
\dfrac{\dot{p}_c}{p_c} = \dfrac{\dot{p}}{p} \left[ 1 + \dfrac{k N \ln R}{M^N} \beta^{N-1} - \dfrac{N \ln R}{M^N} \beta^N \right] = \dfrac{\dot{\beta}}{k - \beta} + \dfrac{k N \ln R}{M^N} \dfrac{\beta^{N-1} \dot{\beta}}{k - \beta} - \dfrac{N \ln R}{M^N} \dfrac{\beta^N \dot{\beta}}{k - \beta} .
\label{eq:casm_pld_p_c_dot_over_p_c}
\end{equation}

From the hardening rule \eqref{eq:hardening_mcc}, $\Delta \varepsilon^p_v$ can be computed as:
\begin{equation}
\Delta \varepsilon^p_v = \dfrac{1}{\theta} \biggl[ - \ln (k-\beta) |_{\beta_n}^{\beta_{n+1}} + \dfrac{k N \ln R}{M^N} \int_{\beta_n}^{\beta_{n+1}} \dfrac{\beta^{N-1}}{k - \beta} d \beta - \dfrac{N \ln R}{M^N} \int_{\beta_n}^{\beta_{n+1}} \dfrac{\beta^N}{k - \beta} d \beta \biggr] .
\label{eq:casm_pld_delta_e_p_v}
\end{equation}

Denote $F_N(x)$ as the primitive integral of $\dfrac{x^N}{1-x}$, such that $F_N'(x) = \dfrac{x^N}{1-x}$, \Cref{eq:casm_pld_delta_e_p_v} can be rewritten as:
\begin{equation}
\Delta \varepsilon^p_v = \dfrac{1}{\theta} \left( - \ln (k-\beta) + \dfrac{k^N N \ln R}{M^N} \left[ F_{N-1} \left( \dfrac{\beta}{k} \right) - F_{N} \left( \dfrac{\beta}{k} \right) \right] \right) \bigg|_{\beta_n}^{\beta_{n+1}} .
\label{eq:casm_pld_d_p_v}
\end{equation}

It is noted that, in the case that $N$ is a positive integer, $F_N(x)$ can be determined recursively as following:
\begin{equation}
F_0(x) = -\ln (1-x) \, , \qquad F_{N+1}(x) = F_N(x) - \dfrac{x^{N+1}}{N+1} .
\end{equation}

\subsection{Solution for deviatoric strain}

\subsubsection{Elastic domain}

The rate of elastic shear strain follows by $\dot{\varepsilon}^e_q = \dfrac{\dot{q}}{3 \mu}$, hence:
\begin{equation}
\dot{\varepsilon}^e_q = \dfrac{\kappa k}{3 r (1+e)} \dfrac{\dot{p}}{p} .
\label{eq:mcc_pld_e_q_dot}
\end{equation}

In the elastic domain: $\dot{\varepsilon}_q = \dot{\varepsilon}^e_q$. Therefore, $\Delta \epsilon_q = \Delta \varepsilon^e_q$ can be approximated by:
\begin{equation}
\Delta \epsilon_q = \dfrac{\kappa k}{3 r (1 + e)} \Delta \left( \ln p \right) = \dfrac{\kappa k}{3 r (1 + e)} \ln \dfrac{p_{n+1}}{p_n} .
\label{eq:casm_pld_d_e_q}
\end{equation}

\subsubsection{Plastic domain}

In the plastic domain, $\Delta \varepsilon^e_q$ can be calculated using \Cref{eq:casm_pld_d_e_q}. To compute $\Delta \varepsilon^^p_q$, we start first by the assumption: 
\begin{equation}
\dot{\varepsilon}^p_q \dfrac{\partial g}{\partial p} = \dot{\varepsilon}^p_v \dfrac{\partial g}{\partial q} \Leftrightarrow \dot{\varepsilon}^p_q = \dot{\varepsilon}^p_v \, \dfrac{3(3 + M) - 2M \beta}{9(M - \beta)} .
\label{eq:casm_pld_d_e_p_q}
\end{equation}

From \Cref{eq:casm_pld_d_e_p_q,eq:casm_pld_p_c_dot_over_p_c}, we come up with the equation for $\dot{\varepsilon}^p_q$ which only depends on $\beta$ and $\dot{\beta}$:
\begin{equation}
\begin{split}
\dot{\varepsilon}^p_q = \dfrac{1}{\theta} \dfrac{3(3 + M) - 2M \beta}{9(M - \beta)} \left( \dfrac{\dot{\beta}}{k - \beta} + \dfrac{k N \ln R}{M^N} \dfrac{\beta^{N-1} \dot{\beta}}{k - \beta} - \dfrac{N \ln R}{M^N} \dfrac{\beta^N \dot{\beta}}{k - \beta} \right) .
\end{split}
\label{eq:casm_pld_d_e_p_q_1}
\end{equation}

Using the relation:
\begin{equation}
\dfrac{3(3+M) - 2M \beta}{(M - \beta)(k - \beta)} = \dfrac{A}{M - \beta} + \dfrac{B}{k - \beta} \, , \qquad A = \dfrac{2M^2 - 3(3+M)}{M-k} \, , \qquad B = \dfrac{3(3+M) - 2kM}{M-k} .
\end{equation}

Then the \Cref{eq:casm_pld_d_e_p_q_1} can be expanded as:
\begin{equation}
\begin{split}
\dot{\varepsilon}^p_q &= \dfrac{A}{9 \theta} \left( \dfrac{\dot{\beta}}{M - \beta} + \dfrac{k N \ln R}{M^N} \dfrac{\beta^{N-1} \dot{\beta}}{M - \beta} - \dfrac{N \ln R}{M^N} \dfrac{\beta^N \dot{\beta}}{M - \beta} \right) + \dfrac{B}{9 \theta} \left( \dfrac{\dot{\beta}}{k - \beta} + \dfrac{k N \ln R}{M^N} \dfrac{\beta^{N-1} \dot{\beta}}{k - \beta} - \dfrac{N \ln R}{M^N} \dfrac{\beta^N \dot{\beta}}{k - \beta} \right) \\
&= \dfrac{A}{9 \theta} \left( \dfrac{\dot{\beta}}{M - \beta} + \dfrac{k N \ln R}{M^N} \dfrac{\beta^{N-1} \dot{\beta}}{M - \beta} - \dfrac{N \ln R}{M^N} \dfrac{\beta^N \dot{\beta}}{M - \beta} \right) + \dfrac{B}{9} \dot{\varepsilon}^p_v .
\end{split}
\end{equation}

Which leads to the equation of $\Delta \varepsilon^p_q$:
\begin{equation}
\Delta \varepsilon^p_q = \dfrac{A}{9 \theta} \left( -\ln (M-\beta) + N \ln R \left[ \dfrac{k}{M} F_{N-1} \left( \dfrac{\beta}{M} \right) - F_N \left( \dfrac{\beta}{M} \right) \right] \right) \bigg|_{\beta_n}^{\beta_{n+1}} + \dfrac{B}{9} \Delta \varepsilon^p_v .
\label{eq:casm_pld_d_p_q}
\end{equation}

%.
%.
%
%\end{verbatim}

\end{document}